\documentclass[
  aps,
  prl,
  reprint,
  superscriptaddress,
  nofootinbib,
  longbibliography
]{revtex4-2}

\usepackage{amsmath,amssymb,mathtools}
\usepackage{bm}
\usepackage{balance}
\usepackage{graphicx}
\usepackage{microtype}
\usepackage{xcolor}
\usepackage{tikz}
\usetikzlibrary{arrows.meta}
\usepackage[colorlinks=true,allcolors=blue!55!black]{hyperref}

\newcommand{\Var}{\operatorname{Var}}
\newcommand{\tr}{\operatorname{tr}}
\newcommand{\ph}{\bm{\phi}}
\newcommand{\cv}{\bm c}
\newcommand{\hv}{\bm h}
\newcommand{\uv}{\bm u}
\newcommand{\nv}{\bm n}
\newcommand{\EE}{\mathbb E}
\newcommand{\PP}{\mathbb P}
\newcommand{\ee}{\bm e}
\newcommand{\F}{\mathrm F}
\newcommand{\op}{\mathrm{op}}
\providecommand{\ket}[1]{\lvert #1\rangle}
\providecommand{\bra}[1]{\langle #1\rvert}

\begin{document}
\title{Representation Learning with Quantum Signal Processing}
\author{Junqi Wang}
\email{junqi.wang@pitt.edu}
\author{Junyu Liu}
\email{junyuliu@pitt.edu}
\affiliation{Department of Computer Science, University of Pittsburgh, Pittsburgh, PA 15260, USA}
\date{November 16, 2026}
\begin{abstract}
Representation learning begins when training changes the features that define similarity between data. A frozen-kernel model only reweights a fixed geometry. We establish quantum signal processing (QSP) as a solvable quantum model of the representation-learning regime. At arbitrary depth, we compute the exact mean and variance of its quantum neural tangent kernel, revealing an input-dependent angular geometry whose diagonal remains non-self-averaging even when the underlying unitary approaches Haar randomness. We also prove a sparse-data theorem for the full nonlinear gradient flow without freezing or ensemble-averaging the kernel: the realized dynamics converges to an integrable scalar flow with a time-dependent kernel closure and explicit convergence times. A finite-depth speed limit holds for every data set and trajectory. At higher data density, numerical results show coupled evolution beyond both the scalar and frozen-kernel descriptions. These results give a controlled theory of learned quantum data geometry with provable training dynamics beyond the frozen limit.
\end{abstract}
\maketitle

\section{Introduction}
 
Learning can adjust both an output function and the representation in which data are compared. For a differentiable model, the tangent feature vector at an input is the gradient of the output with respect to its trainable parameters. Its Gram matrix is the neural tangent kernel (NTK). In an infinite-width or lazy limit these features barely move, the NTK is effectively frozen, and training becomes kernel regression~\cite{Jacot2018,Lee2019,Chizat2019}. Its fixed features prevent the model from changing its own notion of similarity during training.
 
The quantum neural tangent kernel (QNTK) brings this learning-theory viewpoint to variational quantum circuits~\cite{Liu2022,Liu2023,Abedi2023,Shirai2024}. A running QNTK distinguishes representation learning from the frozen regime: when the circuit parameters move, the data-dependent tangent features and their pairwise geometry move with them~\cite{Liu2022}. The same language has clarified the distinction between laziness and barren plateaus, the effect of symmetry on QNTK dimension, and practical QNN diagnostics~\cite{LiuLinJiang2024,WangPruning2023,Scala2026}. Most existing guarantees assume wide, large-qubit, perturbative, or lazy limits~\cite{Liu2023,GarciaMartin2025,Abedi2023}. Here we vary circuit depth instead. In classical networks, a finite depth-to-width ratio already makes the tangent kernel fluctuate and evolve~\cite{HaninNica2020}, and QSP takes the limit of fixed width and growing depth. A deep circuit may randomize its state without concentrating its tangent geometry. This work asks whether a provable training law remains possible when the realized kernel is random at initialization and continues to evolve during learning.
 
Quantum signal processing (QSP) makes this question concrete. QSP is a depth-only, single-qubit polynomial model and the primitive underlying quantum singular-value transformation and many quantum algorithms~\cite{LowChuang2017,LowChuang2019,Gilyen2019,Martyn2021}. Its phases are trainable, while circuit width and Hilbert-space dimension remain fixed; in quantum machine learning, the same structure is a single-qubit data re-uploading model with $Z$-phase processing gates~\cite{PerezSalinas2020,Schuld2021,Yu2022,Barthe2024}. Lin and collaborators made high-degree QSP phase synthesis numerically practical~\cite{Dong2021}, characterized the nonconvex landscape of symmetric phase optimization and proved a locally strongly convex basin~\cite{Wang2022}, and established a controlled $d\to\infty$ limit for phases representing smooth targets through infinite QSP~\cite{DongInfinite2024}. These works study inverse synthesis around structured, target-dependent solutions. Here, ``representation learning'' has the learning-theory meaning: the tangent feature map induced by those phases changes under gradient flow. We initialize all phases independently and uniformly, placing the problem far from a prescribed solution and outside a deterministic frozen-kernel description.
 
This setting exposes two linked difficulties. First, the initial kernel is a quenched random function of the data indices, so replacing it by its ensemble mean can erase fluctuations that survive at large depth. Second, the same realization evolves with the residuals, so even a fixed random-kernel model is insufficient. Architecture-specific quantum kernels already show how physical nonlinearities can modify convergence and generalization~\cite{LiuKerr2023}. Finite-width tangent hierarchies, differential QNTKs, and nonlinear mean-field theories organize corrections beyond frozen kernels, and quantum neural networks can show time-dependent kernels and depth-driven dynamical transitions~\cite{Huang2020,Seleznova2022,Liu2023,You2023,Zhang2024}. QSP complements these approaches with an exactly tractable, realization-resolved limit.
 
At every depth, we solve the initialization geometry: the exact mean kernel is local in the angular data coordinate and its spectrum is Chebyshev structured; at large depth, its normalized diagonal converges to a nondegenerate beta law. For sufficiently sparse angular data, we prove that the full nonlinear QSP gradient flow converges on finite rescaled-time intervals to independent integrable scalar equations. A higher-density numerical example then shows how overlapping tangent neighborhoods produce coupled, time-dependent dynamics beyond both the scalar law and the frozen kernel prediction.
 
\section{QSP as a representation-learning model}
For $x\in[0,1]$, a degree-$d$ QSP sequence and its real response are
\begin{equation}
\begin{aligned}
 U_d(x,\boldsymbol{\phi})&=e^{i\phi_0Z}\prod_{j=1}^{d}\left[W(x)e^{i\phi_jZ}\right],\\
 g_d(x,\boldsymbol{\phi})&=\operatorname{Re}\bra{0}U_d(x,\boldsymbol{\phi})\ket{0},
 \label{eq:model}
\end{aligned}
\end{equation}
where $W(x)=xI+i\sqrt{1-x^2}X$, and $\boldsymbol{\phi}=(\phi_0,\ldots,\phi_d)$ contains $M=d+1$ phases. For fixed phases, $g_d$ extends to a real polynomial of degree at most $d$ with parity $d$. At initialization, all phases are drawn independently and uniformly from $[0,2\pi)$. Given $n$ training pairs $\{(x_a,f_a)\}$, with $x_a\in[0,1]$ and $f_a\in[-1,1]$, we define the residuals and mean-squared loss by $e_a(\boldsymbol{\phi})=g_d(x_a,\boldsymbol{\phi})-f_a$ and $L(\boldsymbol{\phi})=(2n)^{-1}\sum_{a}e_a^2$. The QNTK is defined as
\begin{equation}
 K_{ab}(\boldsymbol{\phi})=\nabla_{\boldsymbol{\phi}}g_d(x_a,\boldsymbol{\phi})\cdot\nabla_{\boldsymbol{\phi}}g_d(x_b,\boldsymbol{\phi}).
 \label{eq:qntk}
\end{equation}
Under continuous gradient flow, $\dot{\boldsymbol{\phi}}=-\nabla_{\boldsymbol{\phi}}L$, the residuals obey the exact dynamics
\begin{equation}
 \dot{\boldsymbol e}=-\frac{1}{n}K(\boldsymbol{\phi}(t))\boldsymbol e .
 \label{eq:dynamics}
\end{equation}
Here $K(\boldsymbol{\phi}(t))$ is the realized, time-dependent QNTK. Define the tangent representation $\boldsymbol\Psi_{\boldsymbol\phi}(x)=\nabla_{\boldsymbol\phi}g_d(x,\boldsymbol\phi)$, with $K_{ab}(\boldsymbol\phi)=\boldsymbol\Psi_{\boldsymbol\phi}(x_a)\cdot\boldsymbol\Psi_{\boldsymbol\phi}(x_b)$. A frozen-kernel model keeps $K$ fixed at $K(\boldsymbol\phi(0))$. QSP representation learning means that $\boldsymbol\Psi_{\boldsymbol\phi(t)}$ and hence the full data-indexed matrix $K_{ab}(t)$ evolve as the phases are trained.
 
\section{Random quantum data geometry}
 
\begin{figure*}[t]
\centering
\includegraphics[width=0.98\textwidth]{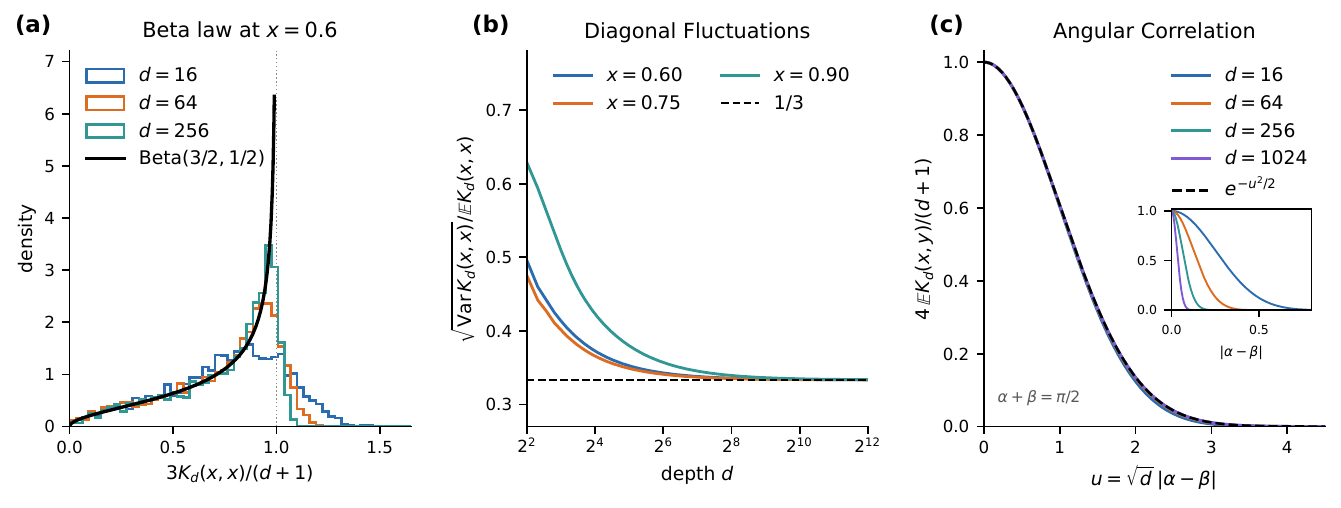}
\caption{\textbf{Non-self-averaging initialization and angular locality of the mean QNTK.} (a) Distribution of the normalized diagonal QNTK $z=3K_d(x,x)/M$ at $x=0.6$. The distributions approach the $\operatorname{Beta}(3/2,1/2)$ law (black). The finite-depth weight above $z=1$ reflects the residual anisotropy of the frame tensor. (b) Exact coefficient of variation $\sqrt{\operatorname{Var}K_d(x,x)}/\mathbb{E}K_d(x,x)$ for three fixed interior inputs. All three curves approach the nonzero asymptotic value $1/3$ (dashed). (c) Along $\alpha+\beta=\pi/2$, the normalized mean kernel $4\mathbb{E}K_d(x,y)/M=\cos^d(\alpha-\beta)$ approaches $e^{-u^2/2}$ under the rescaling $u=\sqrt{d}\,|\alpha-\beta|$.}
\label{fig:initialization}
\end{figure*}

Write $x=\cos\alpha$ and $y=\cos\beta$. Products of QSP responses can be written as transfer processes over copies of the circuit. Averaging each independent phase imposes a matching condition between the $Z$ charges of the copies, reducing the calculation to finite-dimensional balanced subspace. For two copies, this gives the exact finite-depth mean QNTK
\begin{equation}
\mathbb E K_d(x,y)=\frac{M}{4}\left[\cos^d(\alpha-\beta)+\cos^d(\alpha+\beta)\right].
\label{eq:mean}
\end{equation}
The direct term depends on the angular separation $\alpha-\beta$ and measures local similarity, while the reflected term $\alpha+\beta$ enforces the parity and boundary structure of QSP polynomials. Thus the data indices of $K_{ab}$ are not interchangeable: their angular placement determines which residuals can communicate during training. Setting $y=x$ gives $\mathbb E K_d(x,x)=M[1+(2x^2-1)^d]/4$. Four copies, restricted to balanced charge subspaces, give the second moment exactly at every finite depth, in closed form on the diagonal. The full expressions and the detail of transfer matrix methods are given in the Supplemental Material~\cite{SM}. For every fixed interior input $x\in(0,1)$,
\begin{equation}
\begin{aligned}
\frac{\mathbb E K_d(x,x)}{M}&\longrightarrow\frac14,\qquad
\frac{\operatorname{Var}K_d(x,x)}{M^2}\longrightarrow\frac1{144},\\
\frac{\sqrt{\operatorname{Var}K_d(x,x)}}{\mathbb E K_d(x,x)}&\longrightarrow\frac13.
\label{eq:nonconcentration}
\end{aligned}
\end{equation}
Thus both the mean and standard deviation grow linearly with depth, and increasing the number of phases does not suppress the relative fluctuations of a diagonal QNTK entry. The QSP unitary still randomizes completely. For every fixed $x\in(0,1)$, the random walk generated by $W(x)e^{i\phi Z}$ approaches Haar measure on $SU(2)$~\cite{ItoKawada1940,Applebaum2014}. Consequently,
\begin{equation}
\begin{aligned}
&U_d(x,\boldsymbol{\phi})\xRightarrow[d\to\infty]{\mathrm{dist.}}H\sim\operatorname{Haar}(SU(2)),\\
&g_d(x,\boldsymbol{\phi})\xRightarrow[d\to\infty]{\mathrm{dist.}}G\equiv\operatorname{Re}\bra{0}H\ket{0},\\
&p_G(g)=\frac{2}{\pi}\sqrt{1-g^2},\qquad -1<g<1.
\end{aligned}
\label{eq:semicircle}
\end{equation}
To identify the origin of the surviving fluctuations, write $U_d=g_dI+i\boldsymbol c_d\cdot\boldsymbol{\sigma}$. In the frame representation (End Matter), moving the $Z$ rotation of the $j$th phase gate to the left end of $U_d$ turns it into a rotation of the whole circuit about a Bloch-sphere axis $\boldsymbol n_j$, so the derivative of $g_d$ with respect to $\phi_j$ is the projection of $\boldsymbol c_d$ onto this axis. It follows exactly that $K_d(x,x)=\boldsymbol c_d^{\mathsf T}S_d\boldsymbol c_d$, where $S_d=\sum_{j=0}^{d}\boldsymbol n_j\boldsymbol n_j^{\mathsf T}$. At random initialization the axes form a freely rotating chain: consecutive axes enclose the fixed angle $2\alpha$ set by the signal gate, $\boldsymbol n_{j+1}\cdot\boldsymbol n_j=2x^2-1$, while the independent phase of each layer sets the azimuth. The axes are thus a Markov chain that has no memory of its orientation, and at large depth the normalized frame tensor becomes isotropic, $S_d/M\to I_3/3$ in probability. Since $\lVert\boldsymbol c_d\rVert^2=1-g_d^2$, this gives, together with the Haar limit,
\begin{align}
\frac{3K_d(x,x)}{M}-(1-g_d(x)^2)&\xrightarrow[d\to\infty]{\mathrm{Prob.}}0 ,\label{eq:frame-reduction}\\
\frac{3K_d(x,x)}{M}&\xRightarrow[d\to\infty]{\mathrm{dist.}}\operatorname{Beta}\left(\frac32,\frac12\right).\label{eq:beta}
\end{align}
The proof of the isotropization is given in the Supplemental Material~\cite{SM}.
The beta law ties the surviving kernel fluctuations to the realized output amplitude, which is also the quantity moved by learning; the unitary has already mixed to Haar. The finite-depth distribution of the normalized diagonal QNTK and its ratio of standard deviation to mean are shown in Figs.~\ref{fig:initialization}(a) and \ref{fig:initialization}(b), respectively.
 
The mean kernel also determines the scale over which different inputs are coupled at initialization. On the physical half interval $\theta\in[0,\pi/2]$, with normalized measure $d\mu(\theta)=(2/\pi)d\theta$, its integral operator is diagonal in the parity-matched cosine basis, equivalently the Chebyshev basis, with eigenvalues
\begin{equation}
\lambda_m=\frac{M}{2}2^{-d}\binom{d}{(d-m)/2},\qquad m=d,d-2,\ldots\geq0.
\label{eq:mercer}
\end{equation}
Its participation ratio, $r_{\mathrm{part}}\equiv(\sum_m\lambda_m)^2/\sum_m\lambda_m^2$, is asymptotic to $\sqrt{\pi d}/2$. The binomial eigenvalues quantify the initialization bias toward different data variations, the kernel-level counterpart of the vanishing high-frequency components of re-uploading models~\cite{Barthe2024}. This motivates the sample density $w=n/\sqrt d$, which measures how many samples occupy one mean-kernel correlation length. Away from the boundary, $\cos^d(\alpha+\beta)$ is suppressed at large depth, while expanding $\cos^d(\alpha-\beta)$ near $\alpha=\beta$ gives the Gaussian profile shown in Fig.~\ref{fig:initialization}(c), with width $O(d^{-1/2})$. The parameter $w$ is its natural occupancy scale: $w\ll1$ describes isolated tangent neighborhoods, whereas $w\gtrsim1$ allows several data points to share features. The rigorous dynamical limit below requires a stronger condition that also controls the accumulated cross-input terms.
 
\section{Provable representation learning for sparse data}
 
Before any large-depth analysis, unitarity gives $|\partial_{\phi_j}g_d|\le1$, so that $K_{aa}\le M$. Hence $\lambda_{\max}(K)\le\operatorname{tr}K\le nM$, and Eq.~\eqref{eq:dynamics} gives $-\dot L=n^{-2}\boldsymbol e^{\mathsf T}K\boldsymbol e\le2ML$. Every gradient-flow trajectory, at every depth and for every data set, obeys $L(t)\ge L(0)e^{-2Mt}$.
 
The typical loss dynamics also depends on how the kernel is distributed across inputs. We take equally spaced angular nodes on the training interval, $\theta_a=(a+\tfrac12)\pi/(2n)$ and $x_a=\cos\theta_a$, $a=0,\ldots,n-1$. Their angular separation is $O(n^{-1})$. When this separation is large compared with the correlation length $O(d^{-1/2})$, different training inputs are weakly coupled. Let $M=d+1\to\infty$ and allow $n$ to grow with $d$. We require
\begin{equation}
\frac{n^{3/2}[1+\log n]}{\sqrt M}\longrightarrow0 .
\label{eq:sparse-condition}
\end{equation}
We introduce the rescaled time $\tau=Mt/(3n)$. For compactness, write $g_a(\tau)=g_d(x_a,\boldsymbol\phi(3n\tau/M))$ and $K_{aa}(\tau)=K_{aa}(\boldsymbol\phi(3n\tau/M))$ below. The QSP outputs reduce, on every fixed interval $0\leq\tau\leq T$, to the scalar dynamics
\begin{equation}
\frac{d\bar g_a}{d\tau}=-(1-\bar g_a^2)(\bar g_a-f_a),\qquad \bar g_a(0)=g_d(x_a,\boldsymbol\phi(0)).
\label{eq:scalar-flow}
\end{equation}
 
\paragraph{Sparse-data theorem.---}
Under Eq.~\eqref{eq:sparse-condition}, for every fixed $T<\infty$,
\begin{equation}
\sup_{0\leq\tau\leq T}\left[\frac{1}{n}\sum_{a}\left|g_a(\tau)-\bar g_a(\tau)\right|^2\right]^{1/2}\xrightarrow[d\to\infty]{\mathrm{Prob.}}0,
\label{eq:sparse-convergence}
\end{equation}
where $\bar g_a$ is the scalar flow of the same realization and the probability is over the random initial phases. Along the same trajectory, the evolving diagonal kernel obeys
\begin{equation}
\sup_{0\leq\tau\leq T}\left[\frac1n\sum_{a}\left|\frac{3K_{aa}(\tau)}{M}-\left(1-g_a(\tau)^2\right)\right|^2\right]^{1/2}\xrightarrow[d\to\infty]{\mathrm{Prob.}}0.
\label{eq:dynamic-kernel-closure}
\end{equation}

\begin{figure*}[t]
\centering
\includegraphics[width=0.96\textwidth]{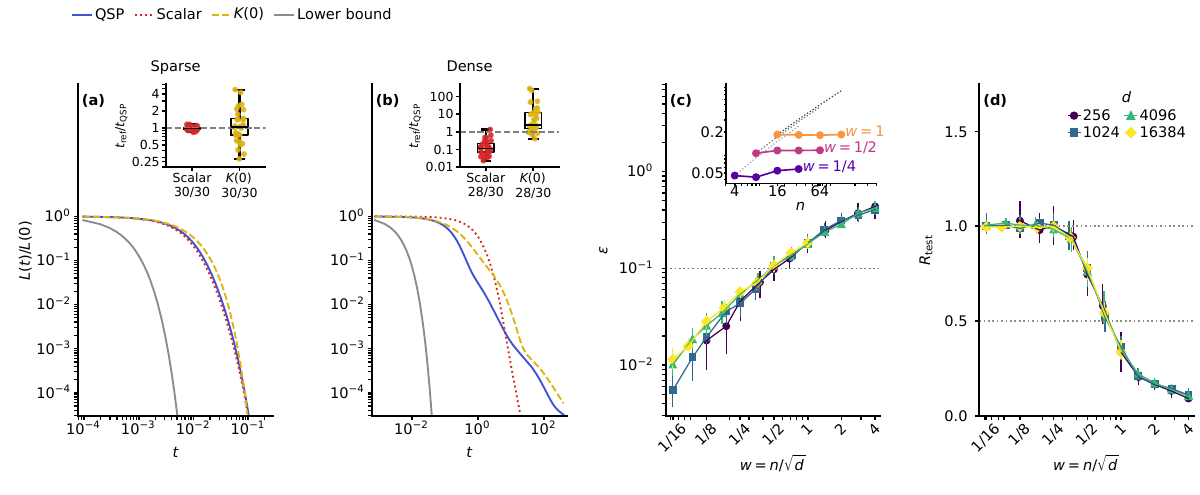}
\caption{\textbf{Learning with isolated and overlapping quantum data neighborhoods.} (a,b) Loss of a representative QSP run compared with the realization-matched scalar flow, the frozen kernel $K(0)$, and the lower bound $e^{-2Mt}$, for (a) $(d,n,w)=(1024,4,0.125)$ and (b) $(128,64,5.66)$. Insets: convergence-time ratios $t_{\mathrm{ref}}/t_{\mathrm{QSP}}$ at $L/L(0)=10^{-4}$ over initializations, with the number of converged runs indicated. (c) Closure error $\epsilon$ and (d) midpoint error ratio $R_{\mathrm{test}}$ versus $w=n/\sqrt d$; points are medians with interquartile bars. Inset of (c): $\epsilon$ versus $n$ at fixed $w$; the dotted curves show the growth $\propto\sqrt n\,(1+\log n)$ allowed by Eq.~\eqref{eq:sparse-condition}.}
\label{fig:dynamics}
\end{figure*}

Here we give an intuitive derivation; the rigorous proof is given in~\cite{SM}. (i)~In the frame representation, Eq.~\eqref{eq:em-anisotropy} splits the output equation into three driving terms. The free part alone gives Eq.~\eqref{eq:scalar-flow}. The self-interaction is set by the anisotropy $B_a=S_{aa}-(M/3)I$, and the coupling by the mixed sums $S_{ab}=\sum_j\nv_{aj}\nv_{bj}^{\mathsf T}$, which measure, layer by layer, how well the rotation axes of two inputs align, and hence how far a phase update driven by input $b$ moves input $a$. The deviation from the scalar flow is a time integral of these terms, so Gr\"onwall's inequality bounds it by a $T$-dependent constant times their largest size on $[0,T]$. (ii)~During training, the axes only rotate, as Eq.~\eqref{eq:em-flow} shows. A common rotation leaves $S_{ab}$ unchanged; only the shear between adjacent axes, of angular velocity $2\phi_j'\nv_{aj}$, changes it. Summation by parts then bounds the growth by the total phase motion, which gradient flow limits to $\sum_j\int_0^T|\phi_j'|\,d\tau\propto\sqrt{nT}$, a factor $\propto\sqrt n$. (iii)~At initialization, the chains of two inputs share the phases but turn by different bond angles, so they stay aligned only over $\xi_{ab}\propto(\theta_a-\theta_b)^{-2}$ layers. The mixed sum $S_{ab}$ is then a random walk of $M/\xi_{ab}$ nearly independent blocks of size $\xi_{ab}$, and Eq.~\eqref{eq:em-cross} gives $\|S_{ab}\|/M\propto\sqrt{\xi_{ab}/M}\propto w/k$ for nodes $k$ spacings apart. Summed over the grid, the coupling is $\propto w(1+\log n)$. The self-interaction follows in the same way from the Markov chain of Eq.~\eqref{eq:em-markov} and is at most $\propto w$. The trajectory error is thus $\propto\sqrt n\,w(1+\log n)\propto n^{3/2}(1+\log n)/\sqrt M$, which vanishes under Eq.~\eqref{eq:sparse-condition}. In typical realizations, the couplings partially cancel, so that $w$ alone sets the closure error, as Fig.~\ref{fig:dynamics}(c) indicates.

In the sparse regime, coupling between inputs becomes subleading, while the diagonal tangent kernel is directly determined by the output, $K_{aa}\simeq M(1-g_a^2)/3$. Unless a sample begins at its target, $g_a$ changes by an order-one amount on the rescaled time scale, so Eq.~\eqref{eq:dynamic-kernel-closure} makes $K_{aa}(t)$ change at leading order as well. The resulting closure is nonlinear and describes an evolving representation even though different inputs become asymptotically independent. For a single input, Eq.~\eqref{eq:scalar-flow} has the form of the asymptotic dynamics of over-parametrized QNNs whose data enter only through the input state; for several samples, their convergence is proved assuming a lower bound on the spectrum of the evolving kernel~\cite{You2023}. In QSP, $W(x)$ enters every layer, so each phase acts along an input-dependent axis, sparse inputs decouple, and no spectral assumption is needed.

We can integrate Eq.~\eqref{eq:scalar-flow} to obtain the explicit solution. For example, when $f_a=0$, define the relative-error convergence time by $\bar g_a(\tau_q)^2=q\bar g_a(0)^2$, with $0<q<1$. One obtains
\begin{equation}
\tau_q=\frac12\log\left[\frac{1-q\bar g_a(0)^2}{q[1-\bar g_a(0)^2]}\right],\qquad t_q=\frac{3n}{M}\tau_q .
\label{eq:convergence-time}
\end{equation}
Because the samples evolve independently, multipoint convergence times follow from the sum of these single-sample losses.
 
\section{Coupled quantum data geometry at higher density}

The scalar law relies on angular locality. The exact cross-input contribution to the output dynamics is $-n^{-1}\sum_{b\neq a}K_{ab}(t)e_b(t)$. When this contribution is no longer subleading, changes at one input modify the outputs at other inputs through the shared tangent geometry, and the dynamics cannot be reduced to independent scalar equations.

Figures~\ref{fig:dynamics}(a) and \ref{fig:dynamics}(b) compare the QSP training trajectory with the scalar flow matched to the same random initialization and with the evolution generated by the frozen kernel $K(0)$. In the sparse case, the scalar flow follows the nonlinear QSP descent. It predicts the convergence time of each of the $30$ runs to within $17\%$, whereas the frozen kernel predicts it up to $3.6$ times too early or $4.8$ times too late, as the inset of Fig.~\ref{fig:dynamics}(a) shows. In the representative high-density case of Fig.~\ref{fig:dynamics}(b), the QSP trajectory departs from both the scalar prediction and the frozen-$K(0)$ reference.

The same QSP model therefore exhibits two distinct forms of representation learning. Figures~\ref{fig:dynamics}(c) and \ref{fig:dynamics}(d) show the typical behavior in a scan trained over the window $0\le\tau\le T$ of Eq.~\eqref{eq:sparse-convergence}, with $T=2$. The closure error $\epsilon$, the left side of Eq.~\eqref{eq:sparse-convergence}, follows a common curve in $w$ for all depths and reaches $0.1$ near $w\approx1/2$. The ratio $R_{\mathrm{test}}$ in panel (d), of the final to the initial root-mean-square residual at the untrained midpoints between adjacent inputs, stays near $1$ for $w\lesssim1/4$, where each input reshapes only its own diagonal kernel entry (the midpoint outputs may still move), and falls below $1/2$ for $w\gtrsim1$, where overlapping inputs share features. In the inset, $\epsilon$ is nearly flat along lines of fixed $w$, far below the growth $\sqrt n\,(1+\log n)$ of the rigorous bound behind Eq.~\eqref{eq:sparse-condition} (fixed $w$ also changes $d$). Thus $w$ sets the crossover.

\section{Discussion and outlook}

QSP separates three notions that are often conflated: randomization, concentration, and representation learning. Increasing QSP depth randomizes the unitary at every fixed interior input. The normalized diagonal QNTK remains realization dependent. The frame picture explains this persistence: the tangent axes become isotropic, while the output amplitude remains random and controls the size of the tangent feature vector. The state approaches Haar randomness while presenting a random metric to the learning problem.

The surviving randomness affects the training trajectory. In the sparse regime, Eqs.~\eqref{eq:sparse-convergence} and \eqref{eq:dynamic-kernel-closure} describe the full QSP learner along individual trajectories. The parameters move, the outputs move, and the diagonal kernel follows them through $K_{aa}(t)\simeq M[1-g_a(t)^2]/3$. The dynamics remains solvable: each datum follows an integrable nonlinear flow with explicit convergence times. This trajectory-level closure establishes provable representation learning beyond the frozen limit. The universal speed limit supplies a complementary statement that holds without sparsity, randomness, or asymptotics.

The exact dependence of $K_{ab}$ on its data indices also suggests a useful theory of quantum data geometry. The QSP encoding turns $x=\cos\theta$ into an angular space with resolution $d^{-1/2}$, Chebyshev eigenmodes, and a depth-dependent spectral bias. The density $w=n/\sqrt d$ then has a direct interpretation. It counts how many observations occupy one tangent neighborhood. Sparse observations primarily change their own diagonal kernel entries. Overlapping observations can reshape shared features and transmit error across the data set. The kernel serves as a convergence-rate matrix and as a dynamical description of how the quantum model organizes data.

The angular geometry organizes the next questions. Varying $w$ locates the crossover from isolated to overlapping tangent neighborhoods in Figs.~\ref{fig:dynamics}(c) and \ref{fig:dynamics}(d); whether the off-diagonal frame sums admit a coupled continuum closure that predicts it remains open. The Chebyshev spectrum provides a mode-resolved route from kernel evolution to generalization, while the scalar sector provides an exact setting in which regularization can be followed through relaxation rates and solution selection. Extending the transport argument to irregular grids and higher-dimensional encodings would determine which aspects of the closure follow from localized quantum feature geometry.

Theorem-level analyses of quantum gradient descent likewise tie controlled learning dynamics to explicit structural hypotheses such as sparsity and dissipation~\cite{LiuLargeScale2024}. Together, these examples place geometry and transport structure at the center of quantum-learning theory.
QSP is a minimal quantum model with a random, time-dependent, and provably tractable learned data geometry. It provides a bridge from frozen quantum kernels to a first-principles theory of nonlinear quantum representation learning.
 
\paragraph*{Data and code availability.---} The code that generates and plots the data in Figs.~\ref{fig:initialization} and \ref{fig:dynamics}, together with the numerical data, is available at \url{https://github.com/AlexWangPhy233/qsp-representation-learning}.
 
\begin{acknowledgments}
\textit{Acknowledgments.---} JW and JL are supported in part by the University of Pittsburgh, School of Computing and Information, Department of Computer Science, Pitt Cyber, Pitt Momentum fund, PQI Community Collaboration Awards, John C. Mascaro Faculty Scholar in Sustainability, Switzerland NSF award 2000-1-243053, NSF award 2535915, 2610010, DOE Genesis Program, Thinking Machines Lab and Cisco Research. This research used resources of the Oak Ridge Leadership Computing Facility, which is a DOE Office of Science User Facility supported under Contract DE-AC05-00OR22725. We acknowledge the use of ChatGPT for discussions and for assistance in writing the numerical code.
\end{acknowledgments}
\balance
\bibliography{references}

\section*{End Matter}

\paragraph*{Frame representation.---}
For an input $x_a=\cos\theta_a$, write $U_a=g_aI+i\cv_a\cdot\bm\sigma$, with $g_a^2+\|\cv_a\|_2^2=1$. Let $U_{a,\le j}$ be the product of the gates up to and including the $j$th phase gate. Moving the $Z$ rotation of that gate to the left end of $U_a$ defines its axis,
\begin{equation}
A_{aj}=U_{a,\le j}\,Z\,U_{a,\le j}^\dagger=\nv_{aj}\cdot\bm\sigma,\qquad \partial_{\phi_j}U_a=iA_{aj}U_a .
\end{equation}
The scalar part of $\partial_{\phi_j}U_a$ gives the exact dictionary
\begin{equation}
\begin{aligned}
&\partial_{\phi_j}g_a=-\nv_{aj}\cdot\cv_a,\qquad K_{ab}=\cv_a^{\mathsf T}S_{ab}\cv_b,\\
&S_{ab}=\textstyle\sum_{j=0}^{d}\nv_{aj}\nv_{bj}^{\mathsf T},
\end{aligned}
\label{eq:em-dictionary}
\end{equation}
where $a,b$ label inputs and each $S_{ab}$ is a $3\times3$ matrix, so that $S=(S_{ab})$ is a matrix-valued matrix indexed by the inputs.

\paragraph*{Dynamics.---}
Under gradient flow, the unitary obeys a Schr\"odinger equation and each axis a Heisenberg equation,
\begin{equation}
\begin{aligned}
&\dot U_a=iH_aU_a,\qquad \dot A_{aj}=i[H_{a,<j},A_{aj}],\\
&H_a=\hv_a\cdot\bm\sigma,\qquad H_{a,<j}=\hv_{a,<j}\cdot\bm\sigma,
\end{aligned}
\label{eq:em-schrodinger}
\end{equation}
with the fields $\hv_{a,<j}=\sum_{k<j}\dot\phi_k\nv_{ak}$ and $\hv_a=\sum_j\dot\phi_j\nv_{aj}=n^{-1}\sum_be_bS_{ab}\cv_b$. The field is the phase velocity mapped into Bloch space and is generated by the residuals. The unitary feels the full field, whereas axis $j$ feels only the earlier phases, so Eq.~\eqref{eq:em-schrodinger} is one closed system rather than two pictures of a single Hamiltonian. In components,
\begin{equation}
\begin{aligned}
&\dot g_a=-\hv_a\cdot\cv_a,\qquad \dot{\cv}_a=g_a\hv_a-\hv_a\times\cv_a,\\
&\dot{\nv}_{aj}=2\nv_{aj}\times\hv_{a,<j}.
\end{aligned}
\label{eq:em-flow}
\end{equation}
These rotations preserve the bond angles $\nv_{aj}\cdot\nv_{a,j+1}=2x_a^2-1$, and the loss obeys $\int_0^t\|\dot{\ph}\|_2^2\,dt'=L(0)-L(t)$. Writing $S_{aa}=(M/3)I+B_a$ splits the output equation, in the rescaled time $\tau=Mt/(3n)$, into
\begin{equation}
\begin{aligned}
g_a'={}&-(1-g_a^2)e_a-\frac{3e_a}{M}\,\cv_a^{\mathsf T}B_a\cv_a\\
&-\frac3M\sum_{b\ne a}e_b\,\cv_a^{\mathsf T}S_{ab}\cv_b,
\end{aligned}
\label{eq:em-anisotropy}
\end{equation}
a free part, which alone gives the scalar flow of Eq.~\eqref{eq:scalar-flow}, a self-interaction set by the anisotropy $B_a$, and a coupling set by the mixed sums $S_{ab}$.

\paragraph*{Markov chain.---}
At initialization, the phase $\phi_j$ commutes with $Z$ and cancels from $\nv_{aj}$, which therefore depends only on $\phi_0,\ldots,\phi_{j-1}$. In the next axis, $\phi_j$ sets the azimuth about $\nv_{aj}$:
\begin{equation}
\nv_{a,j+1}=q_a\nv_{aj}+\sqrt{1-q_a^2}\,\uv_{aj},\qquad q_a=2x_a^2-1,
\label{eq:em-markov}
\end{equation}
where $\uv_{aj}$ is uniform on the unit circle orthogonal to $\nv_{aj}$ and independent of the earlier axes. The axes thus form a time-homogeneous Markov chain on the sphere. Its step multiplies spherical harmonics of degree $\ell$ by $P_\ell(q_a)$. In particular, the anisotropy $X_{aj}=\nv_{aj}\nv_{aj}^{\mathsf T}-I/3$ obeys $\mathbb E[X_{a,j+1}\mid\nv_{a0},\ldots,\nv_{aj}]=\lambda_{2,a}X_{aj}$, with $\lambda_{2,a}=P_2(q_a)$.

\paragraph*{Decay of the interactions.---}
Correlations decay geometrically along the chain, $\mathbb E\operatorname{tr}(X_{aj}X_{ak})=\frac23\lambda_{2,a}^{|j-k|}$, so the self-interaction $B_a=\sum_jX_{aj}$ grows only diffusively, $\mathbb E\|B_a\|_{\mathrm F}^2=O\big(M/(1-\lambda_{2,a})\big)$, and its relative size is $O(M^{-1/2})$. For two inputs, the shared phases make the pair of circuit rotations a joint Markov chain. Its averaged step on paired axes has the gap $(1-\lambda_+)(1-\lambda_-)=2\sin^2(\theta_a-\theta_b)\sin^2(\theta_a+\theta_b)$, which yields~\cite{SM}
\begin{equation}
\begin{aligned}
&\mathbb E\|S_{ab}\|_{\mathrm F}^2\le MA_{ab},\\
&A_{ab}=1+\tfrac12\left[\csc^2(\theta_a-\theta_b)+\csc^2(\theta_a+\theta_b)\right].
\end{aligned}
\label{eq:em-cross}
\end{equation}
Since $|K_{ab}|\le\|S_{ab}\|_{\mathrm F}$, the relative coupling between nodes $k$ spacings apart is at most of order $\sqrt{A_{ab}/M}\sim w/k$. These bounds hold at initialization. During training, the rotations of Eq.~\eqref{eq:em-flow} preserve the norms of $B_a$ and $S_{ab}$ up to the shear between layers, which the loss identity limits~\cite{SM}.

\clearpage
\onecolumngrid
\setcounter{equation}{0}
\setcounter{figure}{0}
\setcounter{section}{0}
\setcounter{secnumdepth}{3}
\renewcommand{\theequation}{S\arabic{equation}}
\renewcommand{\thefigure}{S\arabic{figure}}
\renewcommand{\thesection}{S\arabic{section}}
\renewcommand{\thesubsection}{\thesection.\Alph{subsection}}
\renewcommand{\theHequation}{S\arabic{equation}}
\renewcommand{\theHfigure}{S\arabic{figure}}
\renewcommand{\theHsection}{S\arabic{section}}
\renewcommand{\theHsubsection}{\theHsection.\Alph{subsection}}
\makeatletter
\renewcommand{\p@subsection}{}
\makeatother

\begin{center}
{\large\bfseries Supplemental Material for ``Representation Learning with Quantum Signal Processing''\par}
\end{center}
\vspace{1em}

\section*{Conventions} \label{sec:conventions}
We collect here the conventions used throughout the Supplemental Material.

For $x\in[0,1]$, the QSP sequence is
\begin{equation}
U_d(x,\ph)=e^{i\phi_0Z}\prod_{j=1}^{d} \left[W(x)e^{i\phi_jZ}\right], \qquad W(x)=xI+i\sqrt{1-x^2}\,X, \label{eq:s-model}
\end{equation}
where factors are ordered from left to right with increasing $j$. The number of trainable phases is $M=d+1$, and at initialization $\phi_0,\ldots,\phi_d$ are independent and uniform on $[0,2\pi)$. We write
\begin{equation}
g_d(x,\ph)=\operatorname{Re}\bra{0}U_d(x,\ph)\ket{0}, \qquad J_j(x,\ph)=\partial_{\phi_j}g_d(x,\ph), \qquad K_d(x,y;\ph)=\sum_{j=0}^{d}J_j(x,\ph)J_j(y,\ph). \label{eq:s-kernel}
\end{equation}
For training data $\{(x_a,f_a)\}$,
\begin{equation}
e_a=g_d(x_a,\ph)-f_a, \qquad L(\ph)=\frac1{2n}\sum_{a}e_a^2, \qquad \dot\ph=-\nabla_{\ph}L. \label{eq:s-loss}
\end{equation}
Thus $\dot{\ee}=-K\ee/n$ and $\dot L=-\ee^{\mathsf T}K\ee/n^2$ hold exactly.

\paragraph{Ket and bra copies.---}
Let $V_0=e^{i\phi_0Z}$ and $V_k=W(x)e^{i\phi_kZ}$ for $k\ge1$, so that $U_d=V_0V_1\cdots V_d$, and write $U_{\le j}=V_0\cdots V_j$ and $U_{>j}=V_{j+1}\cdots V_d$. Since $Z$ commutes with $e^{i\phi_jZ}$, differentiating the last factor of $V_j$ gives
\begin{equation}
\partial_{\phi_j}U_d=U_{\le j}(iZ)U_{>j}. \label{eq:s-derivative-insertion}
\end{equation}
Define
\begin{equation}
u(x)=\bra{0}U_d\ket{0},\qquad z_j(x)=\bra{0}U_{\le j}ZU_{>j}\ket{0}. \label{eq:s-copies}
\end{equation}
Taking the real part of the $\bra{0}\cdot\ket{0}$ element of $U_d$ and of Eq.~\eqref{eq:s-derivative-insertion} gives
\begin{equation}
g_d(x)=\tfrac12\bigl[u(x)+\bar u(x)\bigr],\qquad J_j(x)=-\operatorname{Im}z_j(x)=\tfrac{i}{2}\bigl[z_j(x)-\bar z_j(x)\bigr]. \label{eq:s-forward-backward}
\end{equation}
We call $u$ and $z_j$ \emph{ket copies} of the circuit and their complex conjugates \emph{bra copies}. A product of $p$ outputs $g_d$ or derivatives $J_j$ therefore expands into products of $p$ copies, each of which is a ket or a bra.

Writing $x=\cos\theta$, the output extends from the physical interval to a degree-$d$ parity polynomial,
\begin{equation}
g_d(-x,\ph)=(-1)^dg_d(x,\ph), \qquad g_d(\cos\theta,\ph)\in \operatorname{span}\{\cos(d\theta),\cos[(d-2)\theta],\ldots\}. \label{eq:s-parity}
\end{equation}
Unless stated otherwise, all large-$d$ limits hold for fixed $x\in(0,1)$. The endpoint formulas at $x=0$ and $x=1$ are evaluated separately at finite depth.

\section{Transfer matrix method} \label{sec:transfer}

The kernel moments are averages of products of ket and bra copies. This section develops the transfer matrix method for such averages and then applies it to two copies, which give the mean kernel, and to four copies, which give its second moment.

\subsection{Path sums and balanced subspaces} \label{sec:transfer-general}

\paragraph{One copy.---} Let $b_k\in\{0,1\}$ be the $Z$-basis index to the right of $V_k$, and define the $Z$ charge $\chi(0)=+1$, $\chi(1)=-1$. For $k\ge1$, $\bra{b_{k-1}}V_k\ket{b_k}=W(x)_{b_{k-1}b_k}e^{i\phi_k\chi(b_k)}$. Inserting resolutions of the identity between consecutive gates therefore writes a copy as a sum over index paths. Since $V_0$ is diagonal and both ends of the matrix element are $\ket{0}$, every path has $b_0=b_d=0$. An inserted $Z$ is diagonal and sits to the right of $V_j$, so it multiplies the path by $\chi(b_j)$. Hence
\begin{equation}
z_j(x)=\sum_{\substack{b_1,\ldots,b_{d-1}\\ b_0=b_d=0}}\chi(b_j)\left[\prod_{k=1}^{d}W(x)_{b_{k-1}b_k}\right]\exp\!\left[i\sum_{k=0}^{d}\phi_k\chi(b_k)\right], \label{eq:s-path-sum}
\end{equation}
and $u(x)$ is the same sum without the factor $\chi(b_j)$. Each path collects one matrix element per signal gate, one phase factor per phase gate, and one charge per insertion. A bra copy is the same sum with complex-conjugated signal elements and phases.

\paragraph{Several copies.---} Consider $p$ copies $c=1,\ldots,p$ at inputs $x_c$, with $\eta_c=+1$ for a ket and $\eta_c=-1$ for a bra. Collect their indices after the $k$th phase gate into $\bm b_k=(b_k^{(1)},\ldots,b_k^{(p)})\in\{0,1\}^p$, with total charge
\begin{equation}
\chi(\bm b)=\sum_{c=1}^{p}\eta_c\chi\bigl(b^{(c)}\bigr). \label{eq:s-total-charge}
\end{equation}
The product of the copies is a single sum over joint paths,
\begin{equation}
\prod_{c=1}^{p}(\text{copy }c)=\sum_{\substack{\bm b_1,\ldots,\bm b_{d-1}\\ \bm b_0=\bm b_d=\bm0}}\bigl[\text{insertions}\bigr]\left[\prod_{k=1}^{d}\mathcal A_{\bm b_{k-1}\bm b_k}\right]\exp\!\left[i\sum_{k=0}^{d}\phi_k\chi(\bm b_k)\right], \label{eq:s-replica-product}
\end{equation}
where $\mathcal A=\bigotimes_c\mathfrak c_c\bigl(W(x_c)\bigr)$ and $\mathfrak c_c$ complex conjugates bra copies. Each insertion of $Z$ at position $j$ in copy $c$ contributes $\chi(b_j^{(c)})$.

\paragraph{Phase averages.---} The phases are independent, so the average factorizes into one average per phase, and for an integer total charge $\chi$,
\begin{equation}
\EE_{\phi}\,e^{i\phi\chi}=\bm 1\{\chi=0\}. \label{eq:s-phase-matching}
\end{equation}
Every joint index after a phase gate must therefore have zero total charge, and the paths are confined to the balanced subspace
\begin{equation}
\mathcal B=\operatorname{span}\{\,\ket{\bm b}:\chi(\bm b)=0\,\}. \label{eq:s-balanced-general}
\end{equation}
At the two boundaries, all copies carry the fixed index $0$, so $\chi(\bm 0)=\sum_c\eta_c$. The averages of $\phi_0$ and $\phi_d$ therefore remove every product with unequal numbers of kets and bras,
\begin{equation}
\sum_{c=1}^{p}\eta_c=0. \label{eq:s-boundary-condition}
\end{equation}
For interior phases alone this would not be required: two kets, for example, can balance with opposite indices. For $p=2$, Eq.~\eqref{eq:s-boundary-condition} leaves one ket and one bra, and $\dim\mathcal B=2$. For $p=4$, it leaves the six choices of two bras among four copies, each with $\dim\mathcal B=6$.

\paragraph{Transfer matrices.---} Let $P$ project onto $\mathcal B$. A signal gate followed by a phase average acts as the compression
\begin{equation}
T=P\mathcal AP, \label{eq:s-general-transfer}
\end{equation}
and an insertion in the copies $S$ at a given position acts as the diagonal operator $D_S\ket{\bm b}=\prod_{c\in S}\chi(b^{(c)})\ket{\bm b}$. With insertions at ordered positions $j_1\le j_2\le\cdots\le j_r$, the average of Eq.~\eqref{eq:s-replica-product} is the matrix element
\begin{equation}
\EE\prod_{c=1}^{p}(\text{copy }c)=\bra{\bm0}T^{j_1}D_{S_1}T^{j_2-j_1}D_{S_2}\cdots D_{S_r}T^{d-j_r}\ket{\bm0}. \label{eq:s-general-chain}
\end{equation}
We call a chain with insertions \emph{marked}, and the chain without them \emph{unmarked}.

\subsection{Two copies and the mean kernel} \label{sec:transfer-two}

By Eq.~\eqref{eq:s-kernel}, the quantity to compute is
\begin{equation}
\EE K_d(x,y)=\sum_{j=0}^{d}\EE[J_j(x)J_j(y)]. \label{eq:s-mean-target}
\end{equation}

By Eq.~\eqref{eq:s-forward-backward},
\begin{equation}
\EE[J_j(x)J_j(y)]=\tfrac14\EE\bigl[z_j(x)\bar z_j(y)+\bar z_j(x)z_j(y)-z_j(x)z_j(y)-\bar z_j(x)\bar z_j(y)\bigr]. \label{eq:s-two-copy-expansion}
\end{equation}
The last two terms violate Eq.~\eqref{eq:s-boundary-condition} and vanish. The first two are complex conjugates of each other, so $\EE[J_j(x)J_j(y)]=\tfrac12\operatorname{Re}\EE[z_j(x)\bar z_j(y)]$.

\paragraph{Balanced subspace and transfer matrix.---} For a ket at $x$ and a bra at $y$, $\chi(\bm b)=\chi(b)-\chi(b')$, so the two copies carry the same index after every phase gate and $\mathcal B_2=\operatorname{span}\{\ket{00},\ket{11}\}$. A signal gate moves the pair from $\ket{bb}$ to $\ket{cc}$ with amplitude $(T_{xy})_{bc}=W(x)_{bc}\overline{W(y)_{bc}}$. Since $W(x)$ has diagonal elements $x$ and off-diagonal elements $is_x$, with $s_x=\sqrt{1-x^2}$ and similarly $s_y=\sqrt{1-y^2}$, the off-diagonal entries are $(is_x)(-is_y)=s_xs_y$ and
\begin{equation}
T_{xy}=\begin{pmatrix} xy&s_xs_y\\ s_xs_y&xy
\end{pmatrix}.
\label{eq:s-two-transfer}
\end{equation}
The insertion acts on both copies at the same position, $D=\chi(b)\chi(b')$, and equals $1$ on $\mathcal B_2$.

\paragraph{One marked position.---} Equation~\eqref{eq:s-general-chain} therefore gives, for every $j=0,\ldots,d$,
\begin{equation}
\EE[J_j(x)J_j(y)]=\frac12\bra{00}T_{xy}^{j}T_{xy}^{d-j}\ket{00}=\frac12\bra{00}T_{xy}^{d}\ket{00}. \label{eq:s-one-mark}
\end{equation}
The marked position only separates two powers of the same transfer matrix, so the marked chain equals the unmarked one.

\paragraph{Summation and diagonalization.---} For $x=\cos\alpha$ and $y=\cos\beta$, the normalized eigenvectors of $T_{xy}$ are $(1,\pm1)/\sqrt2$, and
\begin{equation}
\operatorname{spec}(T_{xy})= \{xy+s_xs_y,xy-s_xs_y\} =\{\cos(\alpha-\beta),\cos(\alpha+\beta)\}. \label{eq:s-two-spectrum}
\end{equation}
Hence $\bra{00}T_{xy}^d\ket{00}=\tfrac12[\cos^d(\alpha-\beta)+\cos^d(\alpha+\beta)]$. Equation~\eqref{eq:s-one-mark} does not depend on $j$, so the $M=d+1$ terms of Eq.~\eqref{eq:s-mean-target} are equal and
\begin{equation}
\EE K_d(x,y)=\frac{M}{4} \left[\cos^d(\alpha-\beta)+\cos^d(\alpha+\beta)\right]. \label{eq:s-mean-kernel}
\end{equation}
The unmarked chain gives the same matrix element for $\EE[g_d(x)g_d(y)]=\tfrac12\operatorname{Re}\EE[u(x)\bar u(y)]$, so that
\begin{equation}
\EE[g_d(x)g_d(y)]=\frac1M\EE K_d(x,y), \qquad \EE K_d(x,x)=\frac{M}{4}\left[1+(2x^2-1)^d\right]. \label{eq:s-mean-diagonal}
\end{equation}
For fixed interior inputs, both powers in Eq.~\eqref{eq:s-mean-kernel} decay exponentially unless their angular arguments approach a multiple of $\pi$. In the local regime $|\alpha-\beta|=O(d^{-1/2})$,
\begin{equation}
\cos^d(\alpha-\beta)=\exp\!\left[-\frac d2(\alpha-\beta)^2+O\!\left(d|\alpha-\beta|^4\right)\right],
\end{equation}
so the direct term has angular correlation length $O(d^{-1/2})$.

\subsection{Four copies and the variance} \label{sec:transfer-four}

By Eq.~\eqref{eq:s-kernel}, the raw second moment is
\begin{equation}
\EE[K_d(x,y)^2]=\sum_{i,j=0}^{d} \EE[J_i(x)J_i(y)J_j(x)J_j(y)]. \label{eq:s-raw-second-start}
\end{equation}
Order the four copies as $(x,y,x,y)$, so that copies $1,2$ carry the insertion at position $i$ and copies $3,4$ the insertion at position $j$.

\paragraph{Sectors.---} By Eq.~\eqref{eq:s-forward-backward}, each $J$ splits into a ket and a bra copy, and the product of four $J$ expands into $2^4$ terms. Label a term by the set $C\subset\{1,2,3,4\}$ of its bra copies; its prefactor is $(i/2)^4(-1)^{|C|}$. By Eq.~\eqref{eq:s-boundary-condition}, only terms with two bras survive,
\begin{equation}
C\in\{12,13,14,23,24,34\}, \label{eq:s-six-sectors}
\end{equation}
where $12$ abbreviates $\{1,2\}$, and each has prefactor $1/16$. The sectors $C=13,24$ place both bras at the same input and give the two XXYY contractions; the sectors $C=12,14,23,34$ give the four XYXY contractions. In sector $C$, set $\eta_c=-1$ for $c\in C$ and $+1$ otherwise. Equations~\eqref{eq:s-balanced-general} and \eqref{eq:s-general-transfer} give a six-dimensional balanced subspace $\mathcal B_C$ and a transfer matrix $T_C=P_C\mathcal A^{(C)}P_C$.

\paragraph{Insertions.---} The insertions at positions $i$ and $j$ act as
\begin{equation}
D_{12}\ket{\bm b}=\chi(b_1)\chi(b_2)\ket{\bm b}, \qquad D_{34}\ket{\bm b}=\chi(b_3)\chi(b_4)\ket{\bm b}. \label{eq:s-four-insertions}
\end{equation}
For two copies, charge matching forces the two indices to agree, so the insertion is the identity. For four copies it constrains only the signed sum, and $D_{12}$ and $D_{34}$ are nontrivial. On a balanced state, two of the four signed charges $\eta_c\chi(b_c)$ are $+1$ and two are $-1$. Since $\prod_c\eta_c=1$, this implies $\prod_c\chi(b_c)=1$, and hence
\begin{equation}
D_{12}D_{34}=I_{\mathcal B_C}. \label{eq:s-four-insertion-cancellation}
\end{equation}
Since each insertion squares to the identity, we also have $D_{12}=D_{34}$ on $\mathcal B_C$. The insertions therefore cancel when $i=j$, but at distinct positions they are separated by powers of $T_C$ and do not cancel in general.

\paragraph{Marked chains.---} By Eq.~\eqref{eq:s-general-chain}, with the boundary state $\ket{\bm0}=\ket{0000}\in\mathcal B_C$,
\begin{equation}
\EE[K_d(x,y)^2]=\frac1{16}\sum_{|C|=2}\sum_{i,j=0}^{d}\bra{\bm0}\mathcal C_{ij}^{(C)}\ket{\bm0},\qquad
\mathcal C_{ij}^{(C)}=
\begin{cases}
T_C^iD_{12}T_C^{j-i}D_{34}T_C^{d-j},&i\le j,\\
T_C^jD_{34}T_C^{i-j}D_{12}T_C^{d-i},&j<i.
\end{cases}
\label{eq:s-complete-second}
\end{equation}
This expression contains all phase matching and all derivative positions.

\paragraph{Explicit transfer matrices.---} As a representative XYXY sector, take $C=34$ with the ordered balanced basis
\begin{equation}
\ket{00;00},\ket{01;01},\ket{01;10},\ket{10;01},\ket{10;10},\ket{11;11},
\label{eq:s-M6-basis}
\end{equation}
where $\ket{b_1b_2;b_3b_4}$ lists the indices of copies $1$ to $4$, and the semicolon separates the pair at position $i$ from the pair at position $j$. With
\begin{equation}
A=x^2y^2,\quad B=(1-x^2)(1-y^2),\quad \mathcal M=xy\,s_xs_y,\quad L=x^2(1-y^2),\quad R=y^2(1-x^2), \label{eq:s-four-variables}
\end{equation}
the transfer matrix is
\begin{equation}
T_{34}=\begin{pmatrix}
A&L&\mathcal M&\mathcal M&R&B\\
L&A&-\mathcal M&-\mathcal M&B&R\\
\mathcal M&-\mathcal M&A&B&-\mathcal M&\mathcal M\\
\mathcal M&-\mathcal M&B&A&-\mathcal M&\mathcal M\\
R&B&-\mathcal M&-\mathcal M&A&L\\
B&R&\mathcal M&\mathcal M&L&A
\end{pmatrix},
\label{eq:s-M6}
\end{equation}
with the six eigenvalues
\begin{align}
\xi_1&=A-B, & \xi_2&=A-B+L-R=2x^2-1,\nonumber\\
\xi_3&=A+B+L+R=1, & \xi_4&=A-B-L+R=2y^2-1,\nonumber\\
\xi_{5,6}&=\frac12\left[2A+2B-L-R\pm\sqrt{32\mathcal M^2+(L+R)^2}\right]. \label{eq:s-M6-spectrum}
\end{align}
On the diagonal $x=y$, put
\begin{equation}
r_1=2x^2-1, \qquad r_2=6x^4-6x^2+1, \qquad \operatorname{spec}(T_{34})\big|_{x=y}=\{1,1,r_1,r_1,r_1,r_2\}. \label{eq:s-diagonal-M6-spectrum}
\end{equation}
For the XXYY representative $C=13$, use the ordered basis $\ket{0000},\ket{1111},\ket{0011},\ket{0110},\ket{1001},\ket{1100}$. Then
\begin{equation}
T_{13}=\begin{pmatrix}
A&B&\mathcal M&\mathcal M&\mathcal M&\mathcal M\\
B&A&\mathcal M&\mathcal M&\mathcal M&\mathcal M\\
\mathcal M&\mathcal M&A&-L&-R&B\\
\mathcal M&\mathcal M&-L&A&B&-R\\
\mathcal M&\mathcal M&-R&B&A&-L\\
\mathcal M&\mathcal M&B&-R&-L&A
\end{pmatrix},
\label{eq:s-xx-transfer}
\end{equation}
whose spectrum is again Eq.~\eqref{eq:s-M6-spectrum}; only the insertions differ.

\paragraph{Reduction of the six sectors.---} In the bases above, define
\begin{equation}
D_0=\operatorname{diag}(1,-1,-1,-1,-1,1),\qquad
D_1=\operatorname{diag}(1,-1,1,1,-1,1),\qquad
D_2=\operatorname{diag}(1,1,1,-1,-1,1).
\label{eq:s-xy-insertion-representatives}
\end{equation}
The six marked chains reduce to three, with the same insertion at both marked positions:
\begin{equation}
\begin{array}{c|c|c}
C&\text{transfer matrix}&\text{insertion}\\\hline
12,34&T_{34}&D_0\\
14,23&T_{34}&D_1\\
13,24&T_{13}&D_2
\end{array}
\label{eq:s-six-sector-reduction}
\end{equation}
Complementing $C$ complex conjugates the transfer matrix and preserves the basis and the insertions. Balanced states have even bit parity, so every transition contains an even number of off-diagonal signal factors $is$ and the transfer matrix is real. Complementary sectors therefore give identical chains. Exchanging copies $1$ and $3$ maps $C=34$ to $C=14$, preserves the inputs and the boundary state, and maps the insertion $D_0$ to $D_1$.

\paragraph{Double sum.---} If $T_C=\sum_a\xi_aP_a^{(C)}$ is the spectral resolution, the ordering $i\le j$ splits the chain into three segments of lengths $\ell_1=i$, $\ell_2=j-i$, and $\ell_3=d-j$, with $\ell_1+\ell_2+\ell_3=d$. Hence
\begin{equation}
\sum_{0\le i\le j\le d}\bra{\bm0}\mathcal C_{ij}^{(C)}\ket{\bm0}
=\sum_{a,b,c}h_d(\xi_a,\xi_b,\xi_c)\bra{\bm0}P_a^{(C)}D_{12}P_b^{(C)}D_{34}P_c^{(C)}\ket{\bm0},\qquad h_d(a,b,c)=\sum_{\ell_1+\ell_2+\ell_3=d}a^{\ell_1}b^{\ell_2}c^{\ell_3}, \label{eq:s-ordered-spectral-sum}
\end{equation}
where, for distinct arguments,
\begin{equation}
h_d(a,b,c)=\frac{a^{d+2}}{(a-b)(a-c)}+\frac{b^{d+2}}{(b-a)(b-c)}+\frac{c^{d+2}}{(c-a)(c-b)}, \label{eq:s-divided-difference}
\end{equation}
and coincident eigenvalues follow by continuous limits. For the strict ordering $j<i$, the middle segment has positive length, so its spectral factor is $\xi_bh_{d-1}(\xi_a,\xi_b,\xi_c)$, with $D_{12}$ and $D_{34}$ exchanged. This evaluates Eq.~\eqref{eq:s-complete-second} at every finite depth without counting $i=j$ twice. The two unit eigenvalues in Eq.~\eqref{eq:s-diagonal-M6-spectrum} generate a quadratic site sum and hence an $O(d^2)$ raw second moment.

\paragraph{Diagonal second moment.---} For $x=y$, take $1,r_1,r_2$ distinct and let $P_\mu^{(C)}$ project onto the full eigenspace of $T_C$ with eigenvalue $\mu$,
\begin{equation}
P_\mu^{(C)}=\prod_{\substack{\nu\in\{1,r_1,r_2\}\\\nu\ne\mu}}\frac{T_C-\nu I}{\mu-\nu}. \label{eq:s-diagonal-projectors}
\end{equation}
Evaluating the insertion matrix elements and combining the six sectors gives
\begin{equation}
\frac1{16}\sum_{|C|=2}\bra{\bm0}P_\mu^{(C)}D_{12}P_\nu^{(C)}D_{34}P_\kappa^{(C)}\ket{\bm0}
=\delta_{\mu\kappa}w_{\mu\nu},\qquad
(w_{\mu\nu})_{\mu,\nu=(1,r_1,r_2)}
=\frac1{144}\begin{pmatrix}10&0&8\\0&27&0\\8&0&1\end{pmatrix}.
\label{eq:s-diagonal-spectral-weights}
\end{equation}
Only equal outer spectral indices survive the sector sum, so the two outer segments combine to $\mu^{d-\ell}$, and the result depends only on the separation $\ell=|i-j|$:
\begin{equation}
\EE[J_i(x)^2J_j(x)^2]
=\sum_{\mu,\nu}w_{\mu\nu}\mu^{d-\ell}\nu^\ell
=\frac{10+27r_1^d+r_2^d+8(r_2^\ell+r_2^{d-\ell})}{144}.
\label{eq:s-diagonal-marked-correlation}
\end{equation}
This polynomial extends to coincident eigenvalues by continuity. There are $d+1$ terms with $i=j$ and $2(d+1-\ell)$ ordered pairs with separation $\ell\ge1$, and
\begin{equation}
\sum_{\ell=1}^{d}(d+1-\ell)(r_2^\ell+r_2^{d-\ell})=(d+2)\frac{1-r_2^{d+1}}{1-r_2}-(d+1)(1+r_2^d). \label{eq:s-diagonal-geometric-sum}
\end{equation}
With $\Delta_1=r_1^d$, $\Delta_2=r_2^d$, and $1-r_2=6x^2(1-x^2)$, this gives the closed form
\begin{align}
\EE[K_d(x,x)^2]=\frac1{432}\Bigg[&3d^2(27\Delta_1+\Delta_2+10)+6d(27\Delta_1+5\Delta_2+6)\nonumber\\
&+\frac{(8d+16)(\Delta_2-1)}{x^2(x^2-1)}+81\Delta_1+75\Delta_2+6\Bigg], \label{eq:s-diagonal-second}
\end{align}
whose right-hand side is a polynomial after the removable limits are taken. The endpoint branches are
\begin{equation}
\EE[K_d(0,0)^2]=\frac3{16}[1+(-1)^d]M^2, \qquad \EE[K_d(1,1)^2]=\frac38M^2. \label{eq:s-second-endpoints}
\end{equation}
For every fixed $x\in(0,1)$, $|r_1|<1$ and $|r_2|<1$, so
\begin{equation}
\frac{\EE K_d(x,x)}{M}\longrightarrow\frac14, \qquad \frac{\EE[K_d(x,x)^2]}{M^2}\longrightarrow\frac5{72}, \qquad \frac{\Var K_d(x,x)}{M^2}\longrightarrow\frac1{144}, \label{eq:s-moment-limits}
\end{equation}
and
\begin{equation}
\frac{\sqrt{\Var K_d(x,x)}}{\EE K_d(x,x)}=\frac13+\frac{4(3x^4-3x^2+1)}{9x^2(1-x^2)}\frac1d+O(d^{-2}). \label{eq:s-cv-expansion}
\end{equation}
This expansion holds for fixed $x\in(0,1)$ as $d\to\infty$, but not uniformly up to the endpoints.

\section{frame representation, Lax evolution, and the beta law} \label{sec:frame-beta}

\subsection{Exact frame dictionary}

For an input $x_a$, write the $SU(2)$ matrix as
\begin{equation}
U_a=g_aI+i\cv_a\cdot\bm\sigma, \qquad g_a^2+\|\cv_a\|_2^2=1. \label{eq:s-bloch}
\end{equation}
Define the axes $\nv$ and their outer product sums by
\begin{equation}
A_{aj}=U_{a,\le j}ZU_{a,\le j}^{\dagger} =\nv_{aj}\cdot\bm\sigma, \qquad S_{ab}^{<m}=\sum_{j=0}^{m-1}\nv_{aj}\nv_{bj}^{\mathsf T}, \qquad S_{ab}=S_{ab}^{<M}. \label{eq:s-frame-def}
\end{equation}
Every $\nv_{aj}$ is a unit vector, so $\tr S_{aa}=M$. Moving the derivative in Eq.~\eqref{eq:s-derivative-insertion} to the left gives $\partial_{\phi_j}U_a=iA_{aj}U_a$. Using
$(\bm a\cdot\bm\sigma)(\bm b\cdot\bm\sigma)=(\bm a\cdot\bm b)I+i(\bm a\times\bm b)\cdot\bm\sigma$,
we obtain
\begin{equation}
J_j(x_a)=-\nv_{aj}\cdot\cv_a, \qquad K_{ab}=\cv_a^{\mathsf T}S_{ab}\cv_b.\label{eq:s-frame-dictionary}
\end{equation}
The corresponding second derivatives are
\begin{equation}
\partial^2_{\phi_j\phi_l}g_a =-(\nv_{aj}\cdot\nv_{al})g_a +(\nv_{a,\min(j,l)}\times\nv_{a,\max(j,l)})\cdot\cv_a. \label{eq:s-frame-hessian}
\end{equation}

Under gradient flow,
\begin{equation}
\dot\phi_j=\frac1n\sum_{b=1}^{n}e_b (\nv_{bj}\cdot\cv_b). \label{eq:s-phase-flow}
\end{equation}
Introduce the strict-prefix and full fields
\begin{equation}
\hv_{a,<j}=\sum_{k<j}\dot\phi_k\nv_{ak} =\frac1n\sum_b e_bS_{ab}^{<j}\cv_b, \qquad \hv_a=\hv_{a,<M}. \label{eq:s-frame-fields}
\end{equation}
The QSP dynamics can be rewritten in the frame representation:
\begin{equation}
\dot g_a=-\hv_a\cdot\cv_a, \qquad \dot\cv_a=g_a\hv_a-\hv_a\times\cv_a, \qquad \dot\nv_{aj}=2\nv_{aj}\times\hv_{a,<j}.\label{eq:s-exact-frame-flow}
\end{equation}
In matrix form, the last equation is
\begin{equation}
\dot A_{aj}=i[H_{a,<j},A_{aj}], \qquad H_{a,<j}=\hv_{a,<j}\cdot\bm\sigma. \label{eq:s-lax}
\end{equation}
This is a Lax form. Each $A_{aj}$ evolves by conjugation, making its fixed spectrum $\{+1,-1\}$ and the constraint $\|\nv_{aj}\|=1$ manifest. It also makes the exact preservation of adjacent bond angles transparent:
\begin{equation}
\frac{d}{dt}(\nv_{aj}\cdot\nv_{a,j+1})=2(\nv_{aj}\times\hv_{a,<j})\cdot\nv_{a,j+1}+2\nv_{aj}\cdot(\nv_{a,j+1}\times\hv_{a,<j+1}).
\end{equation}
Using $\hv_{a,<j+1}=\hv_{a,<j}+\dot\phi_j\nv_{aj}$, the two terms cancel and hence
\begin{equation}
\frac{d}{dt}(\nv_{aj}\cdot\nv_{a,j+1})=0. \label{eq:fixed-bond-angle}
\end{equation}
Evaluating the conserved inner product at initialization gives
\begin{equation}
\nv_{aj}\cdot\nv_{a,j+1}=\frac12\tr(A_{aj}A_{a,j+1})=2x_a^2-1. \label{eq:bond-angle}
\end{equation}

\subsection{Fixed-input Haar limit}

We show that, at every fixed interior input, the QSP unitary becomes Haar random as the depth grows. For fixed $x\in(0,1)$, the product $U_d=V_0V_1\cdots V_d$ is a random walk on $SU(2)$ whose steps $V_k=W(x)e^{i\phi_kZ}$, $k\ge1$, are independent and identically distributed. The It\^o--Kawada theorem~\cite{ItoKawada1940,Applebaum2014} states that such a random walk on a compact group converges in distribution to the Haar measure if the step law is (i) adapted, meaning that its support is not contained in a proper closed subgroup, and (ii) aperiodic, meaning that its support is not contained in a coset of a proper closed normal subgroup. We verify both conditions.

Let $\mathbb T=\{e^{i\phi Z}:\phi\in[0,2\pi)\}$, a maximal torus of $SU(2)$. The support of one step is the circle $W(x)\mathbb T$.

\emph{Adapted.} Products of support elements and their inverses generate both $\mathbb T$, through $(Wt_1)^{-1}(Wt_2)=t_1^{-1}t_2$, and the conjugate torus $W(x)\mathbb TW(x)^{-1}$, through $(Wt_1)(Wt_2)^{-1}=Wt_1t_2^{-1}W^{-1}$, for $t_1,t_2\in\mathbb T$. On the Bloch sphere, these tori rotate about $\hat{\bm e}_z$ and about its image under a rotation by $2\theta$ about $\hat{\bm e}_x$, where $x=\cos\theta$. At an interior input, $0<2\theta<\pi$, so the two axes are neither parallel nor antiparallel. Their generators and their commutator then span $\mathfrak{su}(2)$, and the closed subgroup generated by the support is all of $SU(2)$.

\emph{Aperiodic.} The proper closed normal subgroups of $SU(2)$ are $\{I\}$ and $\{\pm I\}$. Their cosets contain at most two elements, whereas the support is a circle, so it lies in no such coset.

Both conditions hold, and the theorem gives
\begin{equation}
U_d(x,\ph)\xRightarrow[d\to\infty]{\mathrm{dist.}} H,\qquad H\sim\operatorname{Haar}(SU(2)).
\label{eq:s-haar}
\end{equation}
The first factor $V_0=e^{i\phi_0Z}$ is independent of the others and does not alter the limit, by the invariance of the Haar measure. At the endpoints, the first condition fails: for $x=1$ the support is $\mathbb T$ itself, and for $x=0$ the two tori coincide.

A Haar-random $H\in SU(2)$ has first column $(a,b)^{\mathsf T}$ uniformly distributed on the unit sphere $|a|^2+|b|^2=1$ in $\mathbb C^2$, and the projection of this three-sphere onto one complex coordinate is uniform on the unit disk. Hence $\bra{0}H\ket{0}=a$ is uniform in the unit disk, and
\begin{equation}
g_d(x,\ph)\xRightarrow[d\to\infty]{\mathrm{dist.}} G\equiv\operatorname{Re}\bra{0}H\ket{0}, \qquad p_G(g)=\frac2\pi\sqrt{1-g^2},\quad -1<g<1. \label{eq:s-semicircle}
\end{equation}

\subsection{Frame isotropization and the beta law}

We now show that the normalized diagonal kernel converges to $\operatorname{Beta}(3/2,1/2)$ at every fixed interior input. The frame dictionary reduces the kernel to the frame tensor and the output, the axes form a Markov chain that makes the frame tensor isotropic, and the Haar limit then fixes the distribution of the output.

\paragraph{Reduction to the frame tensor.---} For a real matrix $A$, let $\|A\|_{\F}=[\tr(A^{\mathsf T}A)]^{1/2}$ denote the Frobenius norm and $\|A\|_{\op}=\sup_{\|v\|_2=1}\|Av\|_2$ the operator norm, where $\|v\|_2=(v^{\mathsf T}v)^{1/2}$. By Eq.~\eqref{eq:s-frame-dictionary}, $K_d(x,x)=\cv^{\mathsf T}S_{xx}\cv$ with $\cv^{\mathsf T}\cv=1-g_d(x)^2\le1$. If the frame tensor were isotropic, $S_{xx}=(M/3)I$, this would give exactly $3K_d(x,x)/M=1-g_d(x)^2$, a function of the output alone, whose distribution is fixed by the Haar limit. The deviation from this value is controlled by the anisotropy of $S_{xx}$,
\begin{equation}
\left|\frac{3K_d(x,x)}M-(1-g_d(x)^2)\right|\le3\left\|\frac{S_{xx}}M-\frac I3\right\|_{\op}\le3\left\|\frac{S_{xx}}M-\frac I3\right\|_{\F}. \label{eq:s-beta-contraction}
\end{equation}
It therefore suffices to show that $S_{xx}/M\to I/3$. Since $S_{xx}=\sum_j\nv_j\nv_j^{\mathsf T}$, the difference is a sum of the deviations of the single axes from isotropy,
\begin{equation}
\frac{S_{xx}}M-\frac I3=\frac{B_M}M,\qquad B_m=\sum_{j=0}^{m-1}X_j,\qquad X_j=\nv_j\nv_j^{\mathsf T}-\frac13I.
\label{eq:s-spin-two-def}
\end{equation}
Here $X_j$ measures the anisotropy of the frame at site $j$ and $B_m$ the accumulated anisotropy over the first $m$ sites. Each $X_j$ is of order one, so the task is to show that these deviations decorrelate along the chain.

\paragraph{Markov chain of the axes.---} Set $q=2x^2-1$, so that $\nv_j\cdot\nv_{j+1}=q$ by Eq.~\eqref{eq:bond-angle}. The phase averaged in the transition from $\nv_j$ to $\nv_{j+1}$ is $\phi_j$. In $A_j=U_{\le j}ZU_{\le j}^{\dagger}$, the last phase gate cancels because it commutes with $Z$. Thus the entire axis history through $j$ depends only on $\phi_0,\ldots,\phi_{j-1}$, while $\phi_j$ remains independent and uniform. In $A_{j+1}$, this phase precedes the next signal gate and rotates the new axis around $\nv_j$. Averaging $\phi_j$ therefore makes the azimuth uniform even conditional on the full axis history, and
\begin{equation}
\nv_{j+1}=q\nv_j+\sqrt{1-q^2}\,\uv_j,\qquad \EE[\uv_j\mid\nv_0,\ldots,\nv_j]=0,\qquad \EE[\uv_j\uv_j^{\mathsf T}\mid\nv_0,\ldots,\nv_j]=\frac12\left(I-\nv_j\nv_j^{\mathsf T}\right),
\end{equation}
where $\uv_j$ is uniform on the unit circle orthogonal to $\nv_j$. It follows that
\begin{align}
\EE[\nv_{j+1}\nv_{j+1}^{\mathsf T}\mid\nv_0,\ldots,\nv_j] &=q^2\nv_j\nv_j^{\mathsf T} +\frac{1-q^2}{2}\left(I-\nv_j\nv_j^{\mathsf T}\right) \notag\\
&=\frac13I+\lambda_2\left(\nv_j\nv_j^{\mathsf T}-\frac13I\right),\qquad \lambda_2=\frac{3q^2-1}2=P_2(q)=6x^4-6x^2+1.
\end{align}
One step of the chain keeps the isotropic part and multiplies the deviation from it by $\lambda_2$,
\begin{equation}
\EE[X_{j+1}\mid\nv_0,\ldots,\nv_j]=\lambda_2X_j.
\label{eq:s-spin-two-transfer}
\end{equation}

\paragraph{Isotropization.---} Iterating Eq.~\eqref{eq:s-spin-two-transfer} and using $\tr(X_j^{\mathsf T}X_j)=2/3$ gives
\begin{equation}
\EE\tr(X_j^{\mathsf T}X_k) =\frac23\lambda_2^{|j-k|}.
\label{eq:s-spin-two-correlation}
\end{equation}
Summing the covariance gives the exact finite-$m$ identity
\begin{equation}
\EE\|B_m\|_{\F}^2=\frac23\left[m+2\sum_{r=1}^{m-1}(m-r)\lambda_2^r\right]. \label{eq:s-frame-second}
\end{equation}
For fixed $x\in(0,1)$, $|\lambda_2|<1$, so the right-hand side grows only linearly with $m$. Therefore,
\begin{equation}
\frac{S_{xx}}M=\frac I3+\frac{B_M}{M}\xrightarrow[d\to\infty]{L^2}\frac I3. \label{eq:s-frame-isotropy}
\end{equation}

\paragraph{Beta law.---} Equations~\eqref{eq:s-frame-isotropy} and \eqref{eq:s-beta-contraction} show that the difference on the left of Eq.~\eqref{eq:s-beta-contraction} tends to zero in mean square, and hence in probability. Together with $g_d(x)\Rightarrow G$ from Eq.~\eqref{eq:s-semicircle}, Slutsky's theorem gives
\begin{equation}
\frac{3K_d(x,x)}M\Rightarrow1-G^2.
\end{equation}
Let $Y=1-G^2$. The two inverse branches are $G=\pm\sqrt{1-Y}$, with $|dG/dY|=1/(2\sqrt{1-Y})$. Using $p_G(g)=2\sqrt{1-g^2}/\pi$, we obtain
\begin{equation}
p_Y(y)
=2p_G\bigl(\sqrt{1-y}\bigr)
\left|\frac{d}{dy}\sqrt{1-y}\right|
=\frac2\pi y^{1/2}(1-y)^{-1/2},
\qquad 0<y<1.
\label{eq:s-beta-density}
\end{equation}
This is the density of $\operatorname{Beta}(3/2,1/2)$, and therefore
\begin{equation}
\frac{3K_d(x,x)}M
\xRightarrow[d\to\infty]{\mathrm{dist.}}
\operatorname{Beta}\!\left(\frac32,\frac12\right),
\qquad x\in(0,1)\ \text{fixed}.
\label{eq:s-beta-law}
\end{equation}
Because $0\le K_d/M\le1$, the first two moments also converge, reproducing Eq.~\eqref{eq:s-moment-limits}. At $x=1$,
\[
\frac{K_d}{M}
=\sin^2\left(\sum_j\phi_j\right)
\sim\operatorname{Beta}\!\left(\frac12,\frac12\right)
\]
exactly at every depth. At $x=0$, odd $d$ gives $K_d=0$, whereas even $d$ again gives the $\operatorname{Beta}(1/2,1/2)$ law.
\section{Spectrum of the ensemble-mean kernel} \label{sec:mean-spectrum}

We now regard the ensemble-mean kernel as the kernel of an integral operator and determine its eigenvalues and eigenfunctions. Write $x=\cos\theta$ and work on the physical half interval $\theta\in[0,\pi/2]$ with normalized measure
\begin{equation}
d\mu(\theta)=\frac2\pi d\theta. \label{eq:s-half-measure}
\end{equation}
The ensemble-mean kernel defines a self-adjoint positive finite-rank operator $\overline{\mathcal K}_d$ on $L^2([0,\pi/2],d\mu)$ by
\begin{equation}
(\overline{\mathcal K}_dh)(\alpha)=\int_0^{\pi/2}\EE K_d(\cos\alpha,\cos\beta)h(\beta)d\mu(\beta). \label{eq:s-mean-operator}
\end{equation}
Thus the spectrum considered below is the operator spectrum of $\overline{\mathcal K}_d$. We seek nonzero functions $h$ and scalars $\lambda$ satisfying the eigenvalue equation
\begin{equation}
\overline{\mathcal K}_dh=\lambda h,
\qquad\text{equivalently}\qquad
\int_0^{\pi/2}\EE K_d(\cos\alpha,\cos\beta)h(\beta)d\mu(\beta)=\lambda h(\alpha). \label{eq:s-mean-eigenproblem}
\end{equation}

For the parity fixed by $d$, use the orthonormal basis of $L^2([0,\pi/2],d\mu)$
\begin{equation}
\psi_m(\theta)=\begin{cases}1,&m=0,\\ \sqrt2\cos(m\theta),&m>0,\end{cases}
\qquad m\ge0,\quad m\equiv d\pmod 2. \label{eq:s-cosine-basis}
\end{equation}
Using
\begin{equation}
\cos^d\delta=2^{-d}\sum_{k=0}^{d}\binom dk\cos[(d-2k)\delta] \label{eq:s-binomial-cosine}
\end{equation}
together with
\begin{equation}
\cos m(\alpha-\beta)+\cos m(\alpha+\beta)=2\cos(m\alpha)\cos(m\beta),
\end{equation}
the ensemble-mean kernel has the finite spectral expansion
\begin{equation}
\EE K_d(\cos\alpha,\cos\beta)=\sum_{\substack{0\le m\le d\\m\equiv d\pmod 2}}\lambda_m\psi_m(\alpha)\psi_m(\beta), \label{eq:s-mean-spectral-expansion}
\end{equation}
where
\begin{equation}
\lambda_m=\frac M2\,2^{-d}\binom{d}{(d-m)/2},\qquad m=d,d-2,\ldots\ge0. \label{eq:s-mercer-spectrum}
\end{equation}
Orthogonality therefore gives
\begin{equation}
\overline{\mathcal K}_d\psi_m=\lambda_m\psi_m,\qquad m=d,d-2,\ldots\ge0. \label{eq:s-mean-mode-eigenproblem}
\end{equation}
These are all the nonzero eigenvalues of $\overline{\mathcal K}_d$. In the parity-matched orthonormal basis, the higher-frequency modes $\psi_{d+2},\psi_{d+4},\ldots$ have eigenvalue zero. Equivalently, on the full Hilbert space,
\begin{equation}
\ker\overline{\mathcal K}_d=\operatorname{span}\{\psi_d,\psi_{d-2},\ldots\}^{\perp}. \label{eq:s-mean-nullspace}
\end{equation}
The operator consequently has rank
\begin{equation}
D=\left\lfloor\frac d2\right\rfloor+1.
\end{equation}

To quantify how many of the $D$ nonzero eigenvalues carry appreciable spectral weight, define the participation ratio
\begin{equation}
r_{\mathrm{part}}=\frac{(\sum_m\lambda_m)^2}{\sum_m\lambda_m^2}, \label{eq:s-participation}
\end{equation}
where both sums run over the nonzero spectrum in Eq.~\eqref{eq:s-mercer-spectrum}. Let
\begin{equation}
p_c=2^{-d}\binom d{d/2}\quad\text{for even }d,\qquad q_c=4^{-d}\binom{2d}{d}.
\end{equation}
Using $\sum_{k=0}^{d}\binom dk^2=\binom{2d}{d}$ gives
\begin{equation}
r_{\mathrm{part}}=
\begin{cases}
\displaystyle\frac{4^d}{2\binom{2d}{d}},&d\ \text{odd},\\[2mm]
\displaystyle\frac{(1+p_c)^2}{2(q_c+p_c^2)},&d\ \text{even}.
\end{cases}
\label{eq:s-participation-exact}
\end{equation}
Stirling's formula $k!\sim\sqrt{2\pi k}(k/e)^k$ yields
\begin{equation}
r_{\mathrm{part}}\sim\frac{\sqrt{\pi d}}2. \label{eq:s-participation-asymptotic}
\end{equation}
Thus, although the ensemble-mean integral operator has exact rank $D=O(d)$, its binomial spectral weight is concentrated on only $O(\sqrt d)$ eigenfunctions. This effective spectral bandwidth is the operator counterpart of the $O(d^{-1/2})$ angular correlation length and motivates the sample density $w=n/\sqrt d$.

\section{Sparse-data theorem and convergence times} \label{sec:sparse}

This section first collects the scalar flow and its convergence times (Sec.~\ref{sec:sparse-times}) and states the sparse-data theorem with its consequence for convergence times (Sec.~\ref{sec:sparse-theorem}). The proof follows in Secs.~\ref{sec:sparse-reduction}--\ref{sec:sparse-conclusion}.

\subsection{Scalar flow and convergence times} \label{sec:sparse-times}

Use the half-Chebyshev midpoint grid
\begin{equation}
\theta_a=\frac{(2a+1)\pi}{4n}, \qquad x_a=\cos\theta_a, \qquad a=0,\ldots,n-1, \label{eq:s-midpoint-grid}
\end{equation}
and the rescaled time $\tau=Mt/(3n)$, with primes denoting $d/d\tau$. The scalar flow is
\begin{equation}
\frac{d\bar g_a}{d\tau} =-(1-\bar g_a^2)(\bar g_a-f_a), \qquad \bar g_a(0)=g_d(x_a,\ph(0)). \label{eq:s-scalar-ode}
\end{equation}

\paragraph{Solutions of the scalar flow.---} For $|f|<1$, separation of variables gives the implicit solution
\begin{equation}
F_f(\bar g(\tau))-F_f(\bar g(0))=-\tau,\qquad F_f(u)=-\frac{\log(1-u)}{2(1-f)} -\frac{\log(1+u)}{2(1+f)} +\frac{\log|u-f|}{1-f^2}. \label{eq:s-scalar-implicit}
\end{equation}
For $f=0$,
\begin{equation}
\bar g(\tau)=\frac{g_0} {\sqrt{g_0^2+(1-g_0^2)e^{2\tau}}},\qquad \tau_q=\frac12\log\frac{1-qg_0^2}{q(1-g_0^2)},\qquad t_q=\frac{3n}{M}\tau_q, \label{eq:s-zero-convergence-time}
\end{equation}
where $\tau_q$ is the time at which the squared residual has decreased by the factor $q\in(0,1)$. For an endpoint target $f=\pm1$, the variable $z=(1+f\bar g)/(1-f\bar g)$ obeys $z+\log z=z_0+\log z_0+4\tau$, so the residual decays algebraically.

\paragraph{Several inputs.---} Let $\overline L=(2n)^{-1}\sum_a(\bar g_a-f_a)^2$. If
\begin{equation}
\max_a\max\{|\bar g_a(0)|,|f_a|\}\le m_*<1, \label{eq:s-interior-control}
\end{equation}
each trajectory stays between its initial value and target, $-2\overline L\le\overline L'\le-2(1-m_*^2)\overline L$, and the time at which $\overline L$ reaches $q\overline L(0)$ obeys
\begin{equation}
\frac12\log\frac1q\le\tau_q \le\frac1{2(1-m_*^2)}\log\frac1q. \label{eq:s-scalar-convergence-bounds}
\end{equation}
For fixed $n$ and $\|f\|_\infty<1$, Eq.~\eqref{eq:s-interior-control} holds with probability arbitrarily close to one for a suitable fixed $m_*$, because each initial output converges in distribution to the continuous law of Eq.~\eqref{eq:s-semicircle}. On a growing grid, the largest initial output need not stay away from one, and Eq.~\eqref{eq:s-interior-control} becomes an additional hypothesis.

\paragraph{Universal speed limit.---} Independently of sparsity, $|J_j|\le1$ by unitarity, so $K_{aa}\le M$ and $\lambda_{\max}(K)\le\tr K\le nM$. The exact loss identity then gives $-\dot L\le2ML$, and every trajectory obeys
\begin{equation}
L(t)\ge L(0)e^{-2Mt},\qquad t_q\ge\frac1{2M}\log\frac1q. \label{eq:s-speed-limit}
\end{equation}

\subsection{Sparse-data theorem} \label{sec:sparse-theorem}

The theorem of the main text compares the full QSP flow with the scalar flow of the same realization. It requires
\begin{equation}
\frac{n^{3/2}(1+\log n)}{\sqrt M}\longrightarrow0. \label{eq:s-sparse-condition}
\end{equation}

\paragraph{Theorem.---} Let $M=d+1\to\infty$. Use the grid in Eq.~\eqref{eq:s-midpoint-grid}, independent uniform initial phases, deterministic targets with $\|f\|_\infty\le1$, and exact continuous gradient flow. If Eq.~\eqref{eq:s-sparse-condition} holds, then for every fixed $T<\infty$,
\begin{equation}
\sup_{0\le\tau\le T}\frac{\|\bm g^{\mathrm{full}}(\tau)-\bar{\bm g}(\tau)\|_2}{\sqrt n}\xrightarrow[M\to\infty]{\mathrm{Prob.}}0. \label{eq:s-sparse-theorem}
\end{equation}
Along the same trajectories, the evolving diagonal kernel obeys
\begin{equation}
\sup_{0\le\tau\le T}\left[\frac1n\sum_{a}\left|\frac{3K_{aa}(\tau)}{M}-(1-g_a(\tau)^2)\right|^2\right]^{1/2}\xrightarrow[M\to\infty]{\mathrm{Prob.}}0. \label{eq:s-dynamic-kernel-closure}
\end{equation}
The leading diagonal of the tangent kernel therefore changes with the learned output throughout the window.

Condition~\eqref{eq:s-sparse-condition} is sufficient, though not known to be necessary: its logarithm comes from the coherent grid sum in Eq.~\eqref{eq:s-harmonic-bound}, and its factor $\sqrt n$ from the worst-case phase budget in Eq.~\eqref{eq:s-omega-bounds}.

\paragraph{Convergence times of the full flow.---} Let $\epsilon_T$ be the left side of Eq.~\eqref{eq:s-sparse-theorem}. Since outputs and targets lie in $[-1,1]$, the relative losses differ by at most $2\epsilon_T/L(0)$. Suppose the scalar relative loss crosses $q$ at $\tau_q^{\mathrm{sc}}<T$ with slope at most $-\nu_q<0$ in a neighborhood. Then for small $\epsilon_T$ the full loss crosses in the same neighborhood, and
\begin{equation}
|\tau_q^{\mathrm{full}}-\tau_q^{\mathrm{sc}}|\le\frac{2\epsilon_T}{\nu_qL(0)}. \label{eq:s-time-transfer}
\end{equation}
Under Eq.~\eqref{eq:s-interior-control}, Eq.~\eqref{eq:s-scalar-convergence-bounds} places $\tau_q^{\mathrm{sc}}$ in a fixed interval, and the slope condition holds for any $\nu_q<2(1-m_*^2)q$. For fixed $n$, the initial loss converges in distribution to a law without an atom at zero, so it stays bounded away from zero in probability. For fixed $n$ and $\|f\|_\infty<1$, choosing $m_*$ and then $T$ as above and letting the excluded probability vanish gives $|\tau_q^{\mathrm{full}}-\tau_q^{\mathrm{sc}}|\to0$ in probability. Multiplication by $3n/M$ converts these times to physical time.

\subsection{Proof, step 1: reduction to the interactions} \label{sec:sparse-reduction}

The proof follows Fig.~\ref{fig:s-proof-map}. The output equation contains the scalar flow and two interactions, the self-interaction of each input and the coupling between inputs. Their time integral drives the deviation from the scalar flow, so the theorem reduces to bounding the interactions on the whole window. Training only rotates the axis chains, which amplifies the initial interactions by at most a factor $1+\Gamma_T$ fixed by the phase motion that gradient flow allows (Sec.~\ref{sec:sparse-dynamics}). The initial interactions are controlled by the Markov chain of the axes and by a transfer matrix for two axis chains (Sec.~\ref{sec:sparse-initial}).

\begin{figure}[h]
\centering
\begin{tikzpicture}[
  box/.style={draw, rounded corners=3pt, align=center, inner sep=5pt, font=\small, fill=#1},
  box/.default={white},
  arr/.style={-{Stealth[length=2.2mm]}, thick}]
\node[box=gray!8] (eq) at (0,0) {\textbf{Driving equation}, Eq.~\eqref{eq:s-output-decomposition}\\[1pt]
  $g_a'=-(1-g_a^2)e_a+\rho_a^{\mathrm{diag}}+\rho_a^{\mathrm{off}}$};
\node[box=gray!8] (gr) at (0,-1.75) {\textbf{Gr\"onwall estimate}, Eq.~\eqref{eq:s-finite-certificate}\\[1pt]
  error $\le Te^{\Lambda T}\sup_{\tau\le T}\|\bm\rho(\tau)\|_2/\sqrt n$};
\node[box=blue!6] (dy) at (0,-3.65) {\textbf{Dynamics of the interactions}, Sec.~\ref{sec:sparse-dynamics}\\[1pt]
  axes only rotate; phase budget and summation by parts\\
  $\sup_\tau(\text{interaction})\le(1+\Gamma_T)\times(\text{initial})$,\quad $\Gamma_T=4\sqrt{3nTL(0)}$};
\node[box=green!7] (s0) at (-3.9,-6.0) {\textbf{Self-interaction}, Sec.~\ref{sec:sparse-initial}\\[1pt]
  Markov chain of axes,\\ Doob maximal inequality\\ $s_0\lesssim n/\sqrt M$};
\node[box=green!7] (c0) at (3.9,-6.0) {\textbf{Coupling}, Sec.~\ref{sec:sparse-initial}\\[1pt]
  paired transfer matrix\\ $\mathfrak C_{\mathrm{init}}\lesssim n(1+\log n)/\sqrt M$};
\node[box=orange!8] (th) at (0,-8.15) {\textbf{Conclusion}, Sec.~\ref{sec:sparse-conclusion}\\[1pt]
  error $\lesssim Te^{\Lambda T}\sqrt n\,n(1+\log n)/\sqrt M\to0$ under Eq.~\eqref{eq:s-sparse-condition}};
\node[box=orange!8] (ti) at (0,-9.75) {\textbf{Theorem and convergence times}, Sec.~\ref{sec:sparse-theorem}};
\draw[arr] (eq) -- (gr);
\draw[arr] (gr) -- (dy);
\draw[arr] (s0.north) -- node[left, font=\footnotesize, xshift=-2pt]{initial size} (dy.south west);
\draw[arr] (c0.north) -- node[right, font=\footnotesize, xshift=2pt]{initial size} (dy.south east);
\draw[arr] (dy) -- (th);
\draw[arr] (th) -- (ti);
\end{tikzpicture}
\caption{Structure of the proof of the sparse-data theorem.}
\label{fig:s-proof-map}
\end{figure}

\paragraph{Driving equation.---} Attach input and time indices to the frame quantities of Eq.~\eqref{eq:s-spin-two-def},
\begin{equation}
X_{a,j}(\tau)=\nv_{aj}(\tau)\nv_{aj}(\tau)^{\mathsf T}-\frac13I, \qquad B_{a,m}(\tau)=\sum_{j<m}X_{a,j}(\tau), \qquad S_{ab}^{<m}(\tau)=\sum_{j<m}\nv_{aj}(\tau)\nv_{bj}(\tau)^{\mathsf T}, \label{eq:s-grid-self-def}
\end{equation}
with $S_{ab}=S_{ab}^{<M}$. Substituting $S_{aa}=MI/3+B_{a,M}$ into the exact frame flow, Eq.~\eqref{eq:s-exact-frame-flow}, gives
\begin{align}
g_a'&=-(1-g_a^2)e_a+\rho_a^{\mathrm{diag}}+\rho_a^{\mathrm{off}}, \label{eq:s-output-decomposition}\\
\rho_a^{\mathrm{diag}}&=-\frac{3e_a}{M}\cv_a^{\mathsf T}B_{a,M}\cv_a, \qquad \rho_a^{\mathrm{off}}=-\frac3M\sum_{b\ne a}e_b\cv_a^{\mathsf T}S_{ab}\cv_b. \label{eq:s-remainders}
\end{align}
The first term is the scalar flow. The second is the self-interaction of input $a$, and the third is its coupling to the other inputs.

\paragraph{Gr\"onwall estimate.---} We compare the two flows with Gr\"onwall's inequality in integral form. If a continuous function $u\ge0$ obeys $u(\tau)\le\alpha(\tau)+\Lambda\int_0^\tau u(s)\,ds$ on $[0,T]$, with $\alpha$ nondecreasing and $\Lambda\ge0$, then $u(\tau)\le\alpha(\tau)e^{\Lambda\tau}$. Let $\bar g_a$ start from the same outputs as the full flow, and let $\bm\delta=\bm g^{\mathrm{full}}-\bar{\bm g}$. Unitarity keeps the full outputs in $[-1,1]$, and the scalar flow has equilibria at $\pm1$, so both trajectories stay in this interval, where $F_a(u)=-(1-u^2)(u-f_a)$ obeys $|F_a'(u)|\le2+2\|f\|_\infty\equiv\Lambda\le4$. Subtracting Eqs.~\eqref{eq:s-scalar-ode} and \eqref{eq:s-output-decomposition} gives $\bm\delta'=\bm F(\bm g^{\mathrm{full}})-\bm F(\bar{\bm g})+\bm\rho$, with $\bm\delta(0)=0$ and $\bm\rho=\bm\rho^{\mathrm{diag}}+\bm\rho^{\mathrm{off}}$. Integrating and using the Lipschitz bound,
\begin{equation*}
\|\bm\delta(\tau)\|_2\le\int_0^\tau\|\bm\rho(s)\|_2\,ds+\Lambda\int_0^\tau\|\bm\delta(s)\|_2\,ds .
\end{equation*}
Gr\"onwall's inequality with $\alpha(\tau)=\int_0^\tau\|\bm\rho(s)\|_2\,ds\le\tau\sup_{s\le\tau}\|\bm\rho(s)\|_2$ then gives
\begin{equation}
\sup_{0\le\tau\le T}\frac{\|\bm g^{\mathrm{full}}(\tau)-\bar{\bm g}(\tau)\|_2}{\sqrt n}\le Te^{\Lambda T}\sup_{0\le\tau\le T}\frac{\|\bm\rho(\tau)\|_2}{\sqrt n}. \label{eq:s-finite-certificate}
\end{equation}
For a fixed window, the trajectory error is therefore a constant times the largest interaction on $[0,T]$.

\paragraph{Interaction strengths.---} Since $\|\cv_a\|_2\le1$ and $|e_a|\le1+\|f\|_\infty$, the two interactions are bounded by
\begin{equation}
s(\tau)=\max_{a,\,m\le M}\frac{\|B_{a,m}(\tau)\|_{\F}}{M},\qquad Q_{a,m}(\tau)=\sum_{b\ne a}\|S_{ab}^{<m}(\tau)\|_{\op}, \label{eq:s-interaction-measures}
\end{equation}
namely
\begin{equation}
\frac{\|\bm\rho^{\mathrm{diag}}(\tau)\|_2}{\sqrt n}\le3(1+\|f\|_\infty)s(\tau),\qquad \frac{\|\bm\rho^{\mathrm{off}}(\tau)\|_2}{\sqrt n}\le(1+\|f\|_\infty)\,\mathfrak C_{\mathrm{full}}(\tau),\qquad \mathfrak C_{\mathrm{full}}^2=\frac9{M^2n}\sum_aQ_{a,M}^2. \label{eq:s-source-bounds}
\end{equation}
The strengths include all prefixes $m<M$, because the transport estimate in Sec.~\ref{sec:sparse-dynamics} expresses each full sum through the prefixes at initialization. For the coupling we also use the prefix average
\begin{equation}
\mathfrak C_{\mathrm{pre}}^2=\frac9{M^3n}\sum_a\sum_{m<M}Q_{a,m}^2,\qquad \mathfrak C_{\mathrm{init}}^2=\mathfrak C_{\mathrm{pre}}(0)^2+\mathfrak C_{\mathrm{full}}(0)^2. \label{eq:s-cross-fields}
\end{equation}

\subsection{Proof, step 2: dynamics of the interactions} \label{sec:sparse-dynamics}

\paragraph{Phase budget.---} Gradient flow gives $L'=-(M/3n)\|\ph'\|_2^2$, so
\begin{equation}
\int_0^T\|\ph'\|_2^2\,d\tau=\frac{3n}{M}[L(0)-L(T)]. \label{eq:s-phase-energy}
\end{equation}
Define $\omega_j(\tau)=4\int_0^\tau|\phi_j'|\,ds$. Cauchy--Schwarz in the phases and in time gives, for $\tau\le T$,
\begin{equation}
\|\bm\omega(\tau)\|_1\le\Gamma_T,\qquad \sqrt M\|\bm\omega(\tau)\|_2\le\Gamma_T,\qquad \Gamma_T=4\sqrt{3nTL(0)}\le4\sqrt{6nT}. \label{eq:s-omega-bounds}
\end{equation}
The total phase motion is thus independent of the depth: more layers share the same budget.

\paragraph{Rotations of the axes.---} By Eq.~\eqref{eq:s-lax}, $\nv_{aj}(\tau)=R_{aj}(\tau)\nv_{aj}(0)$, where $R_{aj}\in SO(3)$ solves $R_{aj}'=\Omega_{aj}R_{aj}$, $R_{aj}(0)=I$, with $\Omega_{aj}v=-2(\sum_{k<j}\phi_k'\nv_{ak})\times v$. Adjacent generators differ by one phase channel, $\Omega_{a,j+1}-\Omega_{aj}=-2\phi_j'\nv_{aj}\times$. Variation of constants and orthogonality then give
\begin{equation}
\|R_{a,j+1}(\tau)-R_{aj}(\tau)\|_{\op}\le2\int_0^{\tau}|\phi_j'|\,ds=\tfrac12\omega_j(\tau). \label{eq:s-adjacent-rotation}
\end{equation}
A common rotation of all layers would leave every interaction unchanged. Only this relative rotation of adjacent layers can change them.

\paragraph{Summation by parts.---} Let $\mathcal L_j$ be linear maps that preserve a norm $\|\cdot\|$ and satisfy $\|(\mathcal L_j-\mathcal L_{j+1})X\|\le\omega_j\|X\|$. For a sequence $X_j$ with prefixes $A_m=\sum_{j<m}X_j$,
\begin{equation}
\sum_{j<m}\mathcal L_jX_j=\mathcal L_{m-1}A_m+\sum_{j<m-1}(\mathcal L_j-\mathcal L_{j+1})A_{j+1},\qquad \Bigl\|\sum_{j<m}\mathcal L_jX_j\Bigr\|\le\|A_m\|+\sum_{j<m-1}\omega_j\|A_{j+1}\|. \label{eq:s-abel}
\end{equation}
For the self-interaction, take $\mathcal L_j(X)=R_{aj}XR_{aj}^{\mathsf T}$ with the Frobenius norm. For the coupling, take $\mathcal L_j(X)=R_{aj}XR_{bj}^{\mathsf T}$ with the operator norm. Both satisfy the hypothesis by Eq.~\eqref{eq:s-adjacent-rotation}, and both map the initial summands to the summands at time $\tau$. With Eq.~\eqref{eq:s-omega-bounds}, the self-interaction obeys
\begin{equation}
s(\tau)\le(1+\Gamma_T)\,s(0). \label{eq:s-self-transport}
\end{equation}
For the coupling, Eq.~\eqref{eq:s-abel} gives $Q_{a,M}(\tau)\le Q_{a,M}(0)+\sum_j\omega_jQ_{a,j+1}(0)$. Cauchy--Schwarz in $j$ and then in $a$, together with $\sqrt M\|\bm\omega\|_2\le\Gamma_T$, gives
\begin{equation}
\mathfrak C_{\mathrm{full}}(\tau)\le\mathfrak C_{\mathrm{full}}(0)+\Gamma_T\,\mathfrak C_{\mathrm{pre}}(0). \label{eq:s-cross-full-final}
\end{equation}
Equations~\eqref{eq:s-finite-certificate}, \eqref{eq:s-source-bounds}, \eqref{eq:s-self-transport}, and \eqref{eq:s-cross-full-final} combine into
\begin{equation}
\sup_{0\le\tau\le T}\frac{\|\bm g^{\mathrm{full}}(\tau)-\bar{\bm g}(\tau)\|_2}{\sqrt n}\le Te^{\Lambda T}\beta_T,\qquad \beta_T=(1+\|f\|_\infty)\bigl[3(1+\Gamma_T)s_0+\mathfrak C_{\mathrm{full}}(0)+\Gamma_T\mathfrak C_{\mathrm{pre}}(0)\bigr], \label{eq:s-beta-T}
\end{equation}
with $s_0=s(0)$. It remains to bound the initial interactions.

\subsection{Proof, step 3: initial size of the interactions} \label{sec:sparse-initial}

\paragraph{Self-interaction: Markov chain of the axes.---} At initialization the axes of input $a$ form the Markov chain of Sec.~\ref{sec:frame-beta}, with fixed bond angle and uniform azimuth. Let $\lambda_{2,a}=P_2(\cos2\theta_a)$, $\gamma_a=1-\lambda_{2,a}=\tfrac32\sin^2(2\theta_a)$, and $\mathcal F_{a,r}=\sigma(\nv_{a0},\ldots,\nv_{ar})$. By Eq.~\eqref{eq:s-spin-two-transfer}, $D_{a,j}=X_{a,j}-\lambda_{2,a}X_{a,j-1}$ has zero conditional mean given $\mathcal F_{a,j-1}$, so $\mathcal M_{a,r}=\sum_{j=1}^rD_{a,j}$ is a matrix-valued martingale, and
\begin{equation}
\gamma_aB_{a,m}=X_{a,0}-\lambda_{2,a}X_{a,m-1}+\mathcal M_{a,m-1}. \label{eq:s-martingale-decomposition}
\end{equation}
By the conditional Jensen inequality, $\|\mathcal M_{a,r}\|_{\F}$ is a nonnegative submartingale. Doob's $L^2$ maximal inequality, the orthogonality of the increments with $\EE\|D_{a,j}\|_{\F}^2=\tfrac23(1-\lambda_{2,a}^2)$ from Eq.~\eqref{eq:s-spin-two-correlation}, and $\|X_{a,0}-\lambda_{2,a}X_{a,m-1}\|_{\F}^2\le2$ give
\begin{equation}
\EE\max_{m\le M}\|B_{a,m}\|_{\F}^2\le\frac{16}{3}M\frac{1+\lambda_{2,a}}{1-\lambda_{2,a}}+\frac4{(1-\lambda_{2,a})^2}. \label{eq:s-doob}
\end{equation}
The large inverse gaps occur only near the ends of the grid, and the grid sums are exact, $\sum_a\gamma_a^{-1}=2n^2/3$ and $\sum_a\gamma_a^{-2}=4(n^4+2n^2)/27$. A union bound and Markov's inequality then give
\begin{equation}
\PP(s_0>\varepsilon_s)\le\frac1{\varepsilon_s^2}\left[\frac{64}{9}\frac{n^2}{M}-\frac{16}{3}\frac nM+\frac{16}{27}\frac{n^4+2n^2}{M^2}\right]. \label{eq:s-self-tail}
\end{equation}

\paragraph{Coupling: transfer matrix for two axis chains.---} The coupling involves the joint statistics of two axis chains that share the phases. Expanding the Frobenius norm,
\begin{equation}
\|S_{ab}^{<m}\|_{\F}^2=\sum_{j,k<m}(\nv_{aj}\cdot\nv_{ak})(\nv_{bj}\cdot\nv_{bk}),\qquad C_{ab}(r)=\EE\bigl[(\nv_{a,k+r}\cdot\nv_{ak})(\nv_{b,k+r}\cdot\nv_{bk})\bigr], \label{eq:s-pair-correlation}
\end{equation}
so the required moment is a two-chain correlation. We evaluate it with a transfer matrix for the Bloch rotations of the circuit.

\emph{Chain.} Let $\mathcal R(U)\in SO(3)$ denote the Bloch rotation $\bm v\cdot\bm\sigma\mapsto U(\bm v\cdot\bm\sigma)U^\dagger$, so that $\nv_{aj}=\mathcal R(U_{a,\le j})\hat{\bm e}_z$ by Eq.~\eqref{eq:s-frame-def}. The prefix rotation is common to $\nv_{a,k+r}$ and $\nv_{ak}$ and cancels from their inner product,
\begin{equation}
\nv_{a,k+r}\cdot\nv_{ak}=\hat{\bm e}_z^{\mathsf T}\,\mathcal W_a\mathcal R_z(\phi_{k+1})\mathcal W_a\mathcal R_z(\phi_{k+2})\cdots\mathcal R_z(\phi_{k+r-1})\mathcal W_a\,\hat{\bm e}_z, \label{eq:s-pair-chain}
\end{equation}
where $\mathcal W_a=\mathcal R(W(x_a))$, $\mathcal R_z(\phi)=\mathcal R(e^{i\phi Z})$, and the last phase drops because $\mathcal R_z\hat{\bm e}_z=\hat{\bm e}_z$. In the Cartesian Bloch basis $\hat{\bm e}_x,\hat{\bm e}_y,\hat{\bm e}_z$, the signal gate is a rotation about $\hat{\bm e}_x$ by $-2\theta_a$,
\begin{equation}
\mathcal W_a\hat{\bm e}_x=\hat{\bm e}_x,\quad \mathcal W_a\hat{\bm e}_y=q_a\hat{\bm e}_y-\bar q_a\hat{\bm e}_z,\quad \mathcal W_a\hat{\bm e}_z=\bar q_a\hat{\bm e}_y+q_a\hat{\bm e}_z,\qquad q_a=\cos2\theta_a,\quad \bar q_a=\sin2\theta_a, \label{eq:s-signal-bloch-rotation}
\end{equation}
where $q_a=2x_a^2-1$ is the bond-angle cosine of Eq.~\eqref{eq:bond-angle}. The two inputs see the same phases, so the product in Eq.~\eqref{eq:s-pair-correlation} is one matrix element of a chain of paired rotations $\mathcal W_a\otimes\mathcal W_b$ and $\mathcal R_z(\phi)\otimes\mathcal R_z(\phi)$ between $\hat{\bm e}_z\otimes\hat{\bm e}_z$ at both ends. The intermediate phases $\phi_{k+1},\ldots,\phi_{k+r-1}$ are independent, so the average factorizes over layers.

\emph{Charge projection.} In the spherical basis $\hat{\bm e}_\pm=(\hat{\bm e}_x\pm i\hat{\bm e}_y)/\sqrt2$, $\mathcal R_z(\phi)$ multiplies $\hat{\bm e}_\pm$ by $e^{\pm2i\phi}$ and leaves $\hat{\bm e}_z$ unchanged, so the Bloch vectors carry the charges $\pm2$ and $0$. The average of each shared phase therefore keeps only pairs with zero total charge, namely $\hat{\bm e}_z\otimes\hat{\bm e}_z$ and $\hat{\bm e}_\pm\otimes\hat{\bm e}_\mp$. Let $\Pi$ project onto this subspace, which has the real basis
\begin{equation}
E_0=\hat{\bm e}_z\otimes\hat{\bm e}_z,\qquad E_1=\frac{\hat{\bm e}_x\otimes\hat{\bm e}_x+\hat{\bm e}_y\otimes\hat{\bm e}_y}{\sqrt2},\qquad E_2=\frac{\hat{\bm e}_x\otimes\hat{\bm e}_y-\hat{\bm e}_y\otimes\hat{\bm e}_x}{\sqrt2}. \label{eq:s-pair-invariant-basis}
\end{equation}
The boundary vector $E_0$ has zero charge, so $\Pi$ may also be inserted at both ends.

\emph{Transfer matrix.} One averaged step is $\mathcal T_{ab}=\Pi(\mathcal W_a\otimes\mathcal W_b)\Pi$. It leaves $E_2$ decoupled and acts on $\operatorname{span}\{E_0,E_1\}$ as
\begin{equation}
T_{ab}=\begin{pmatrix}q_aq_b&\bar q_a\bar q_b/\sqrt2\\ \bar q_a\bar q_b/\sqrt2&(1+q_aq_b)/2\end{pmatrix}. \label{eq:s-pair-transfer}
\end{equation}
Writing $(T_{ab}^r)_{00}$ for the $E_0E_0$ element of $T_{ab}^r$, Eq.~\eqref{eq:s-pair-chain} gives
\begin{equation}
C_{ab}(r)=(T_{ab}^r)_{00},\qquad \EE\|S_{ab}^{<m}\|_{\F}^2=m+2\sum_{r=1}^{m-1}(m-r)\,(T_{ab}^r)_{00}. \label{eq:s-cross-moment}
\end{equation}

\emph{Bound.} The eigenvalues of $T_{ab}$ satisfy
\begin{equation}
(1-\lambda_+)(1-\lambda_-)=2\sin^2(\theta_a-\theta_b)\sin^2(\theta_a+\theta_b),\qquad \bigl[(I-T_{ab})^{-1}\bigr]_{00}=\frac14\left[\csc^2(\theta_a-\theta_b)+\csc^2(\theta_a+\theta_b)\right]. \label{eq:s-pair-gap}
\end{equation}
For $a\ne b$ on the grid, $|\lambda_\pm|<1$: $T_{ab}$ is a compression of an orthogonal map, Eq.~\eqref{eq:s-pair-gap} excludes $1$, and $-1$ would require $q_aq_b=-1$. Since $T_{ab}$ is symmetric, both spectral weights of its $E_0E_0$ element are nonnegative. Modes with $\lambda\ge0$ contribute at most $m/(1-\lambda)$ to the weighted sum in Eq.~\eqref{eq:s-cross-moment}, and modes with $-1<\lambda<0$ give an alternating sum that is nonpositive. Hence
\begin{equation}
\EE\|S_{ab}^{<m}\|_{\F}^2\le mA_{ab},\qquad A_{ab}=1+\frac12\left[\csc^2(\theta_a-\theta_b)+\csc^2(\theta_a+\theta_b)\right]. \label{eq:s-cross-envelope}
\end{equation}
The correlation length of two chains is thus set by their angular separation. On the grid, $|\theta_a-\theta_b|=|a-b|\pi/(2n)$ and $\sin u\ge2u/\pi$ give harmonic sums,
\begin{equation}
\max_a\sum_{b\ne a}\sqrt{A_{ab}}\le Cn(1+\log n), \label{eq:s-harmonic-bound}
\end{equation}
for an absolute constant $C$. This sum is the origin of the logarithm. Minkowski's inequality and $\|S\|_{\op}\le\|S\|_{\F}$ give $[\EE Q_{a,m}(0)^2]^{1/2}\le C\sqrt m\,n(1+\log n)$, and therefore
\begin{equation}
\EE\mathfrak C_{\mathrm{init}}^2\le C_0\frac{n^2}{M}(1+\log n)^2,\qquad \PP(\mathfrak C_{\mathrm{init}}>\varepsilon_c)\le\frac{C_0n^2(1+\log n)^2}{M\varepsilon_c^2}, \label{eq:s-cross-tail}
\end{equation}
for an absolute constant $C_0$.

\subsection{Proof, step 4: conclusion} \label{sec:sparse-conclusion}

For fixed $T$, $\Gamma_T=O(\sqrt n)$, so $\beta_T\to0$ requires the initial interactions to be $o(n^{-1/2})$. Under Eq.~\eqref{eq:s-sparse-condition}, the initialization scales in Eqs.~\eqref{eq:s-self-tail} and \eqref{eq:s-cross-tail} are $o(n^{-1/2})$. For the $n^4/M^2$ term, this follows from $n^{5/2}/M=n^{-1/2}(n^{3/2}/\sqrt M)^2\to0$. Thresholds $\varepsilon_s,\varepsilon_c$ can therefore be chosen with $\sqrt n\,\varepsilon_s,\sqrt n\,\varepsilon_c\to0$ while the probabilities in Eqs.~\eqref{eq:s-self-tail} and \eqref{eq:s-cross-tail} vanish. On the event $\{s_0\le\varepsilon_s,\ \mathfrak C_{\mathrm{init}}\le\varepsilon_c\}$, whose probability tends to one, Eq.~\eqref{eq:s-beta-T} gives $\beta_T\to0$. This proves Eq.~\eqref{eq:s-sparse-theorem}. For Eq.~\eqref{eq:s-dynamic-kernel-closure}, Eq.~\eqref{eq:s-frame-dictionary} gives $3K_{aa}/M-(1-g_a^2)=(3/M)\cv_a^{\mathsf T}B_{a,M}\cv_a$, whose absolute value is at most $3s(\tau)\le3(1+\Gamma_T)s_0$, which tends to zero in probability because $\sqrt n\,s_0\to0$ and $\Gamma_T=O(\sqrt n)$.

\section{Numerical methods} \label{sec:numerics}

The figures of the main text are generated by the released code at \url{https://github.com/AlexWangPhy233/qsp-representation-learning}, which records all parameters, random seeds, and tolerances.

Figure~\ref{fig:initialization} combines Monte Carlo sampling and exact formulas. Panel~(a) samples $4000$ independent uniform-phase initializations at each depth to obtain the distribution of the diagonal kernel, and panels~(b) and (c) evaluate Eqs.~\eqref{eq:s-mean-diagonal}, \eqref{eq:s-diagonal-second}, and \eqref{eq:s-mean-kernel}.

In Fig.~\ref{fig:dynamics}, the inputs lie on the grid of Eq.~\eqref{eq:s-midpoint-grid}, the phases are initialized uniformly, and the targets are the responses of independent degree-$8$ QSP teachers with the parity of $d$. Panels~(a) and (b) train by gradient descent with step size $0.1/M$ over $30$ initializations. Each run is compared with the scalar flow of Eq.~\eqref{eq:s-scalar-ode} and with the continuous flow of the frozen kernel $K(0)$ of the same initialization, and the curve $e^{-2Mt}$ is the bound for continuous gradient flow. The plotted run has the convergence time closest to the median of the converged runs. In panel~(b), $28$ of the $30$ runs reach the threshold $L/L(0)=10^{-4}$ within the cap of $5\times10^5$ steps. Panels~(c) and (d) integrate the gradient flow in rescaled time by a fourth-order Runge--Kutta method with $\Delta\tau=0.05$, with $32$ to $64$ initializations per point. The closure error $\epsilon$ is evaluated at $T=2$, and $R_{\mathrm{test}}$ on the midpoints between adjacent training nodes.

\end{document}